\documentclass[10pt,journal,compsoc]{IEEEtran}

\ifCLASSOPTIONcompsoc
      \usepackage[nocompress]{cite}
\else
    \usepackage{cite}
\fi

\usepackage[utf8]{inputenc}

\usepackage{color, colortbl, soul}
\usepackage[cmex10]{amsmath}
\usepackage{balance}
\usepackage{makeidx}
\usepackage{amssymb}
\usepackage{mathtools}
\usepackage{stmaryrd}
\usepackage{latexsym}
\usepackage{graphicx}
\usepackage[tight,footnotesize]{subfigure}
\usepackage{comment}
\usepackage{enumerate}
\usepackage{booktabs}
\usepackage{array}
\usepackage{algpseudocode}
\definecolor{Gray}{gray}{0.75}
\definecolor{LGray}{gray}{0.95}

\newlength{\barwidth}
\newcommand{\cut}[1]{}

\makeatletter
\newcommand{\colvect}[2][r]{  \gdef\@VORNE{1}
  \left(\hskip-\arraycolsep    \begin{array}{#1}\vekSp@lten{#2}\end{array}  \hskip-\arraycolsep\right)}

\def\vekSp@lten#1{\xvekSp@lten#1;vekL@stLine;}
\def\vekL@stLine{vekL@stLine}
\def\xvekSp@lten#1;{\def\temp{#1}  \ifx\temp\vekL@stLine
  \else
    \ifnum\@VORNE=1\gdef\@VORNE{0}
    \else\@arraycr\fi    #1    \expandafter\xvekSp@lten
  \fi}
\makeatother

\ifCLASSINFOpdf
            \else
                  \fi

\begin{document}

\title{Social Fingerprinting:\\detection of spambot groups through DNA-inspired behavioral modeling}

\author{Stefano~Cresci, 
        Roberto~Di~Pietro,
        Marinella~Petrocchi,
        Angelo~Spognardi,
        and~Maurizio~Tesconi
        \thanks{S. Cresci, M. Petrocchi, and M. Tesconi are with the Institute of Informatics and Telematics, IIT-CNR, Italy.}
\thanks{R. Di Pietro is with Nokia Bell Labs, Paris-France, the University of Padua, Maths Dept., Italy, and the  Institute of Informatics and Telematics, IIT-CNR, Italy.}
\thanks{A. Spognardi is with DTU Compute, Technical University of Denmark, and the Institute of Informatics and Telematics, IIT-CNR, Italy.}
}
\markboth{IEEE Transactions on Dependable and Secure Computing}{Cresci \MakeLowercase{\textit{et al.}}: Social Fingerprinting: detection of spambot groups through DNA-inspired behavioral modeling}
\IEEEtitleabstractindextext{\begin{abstract}Spambot detection in online social networks is a long-lasting challenge involving the study and design of detection techniques capable of efficiently identifying ever-evolving spammers. Recently, a new wave of \emph{social spambots} has emerged, with advanced human-like characteristics that allow them to go undetected even by current state-of-the-art algorithms.
In this paper, we show that efficient spambots detection can be achieved via an in-depth analysis of their collective behaviors exploiting the {\it digital DNA} technique for modeling the behaviors of social network users. Inspired by its biological counterpart, in the digital DNA representation the behavioral lifetime of a digital account is encoded in a sequence of characters. Then, we define a similarity measure for such digital DNA sequences. We build upon digital DNA and the similarity between groups of users to characterize both genuine accounts and spambots. Leveraging such characterization, we design the {\it Social Fingerprinting} technique, which is able to discriminate among spambots and genuine accounts in both a supervised and an unsupervised fashion. 
 We finally evaluate the effectiveness of Social Fingerprinting and we compare it with three state-of-the-art detection algorithms.
Among the peculiarities of our approach is the possibility to apply off-the-shelf DNA analysis techniques to study online users behaviors and to efficiently rely on a limited number of lightweight account characteristics.
\end{abstract}
}

\maketitle
\IEEEdisplaynontitleabstractindextext
\IEEEpeerreviewmaketitle

\makeatletter{}
\section{Introduction}
\label{sec:introduction}

Online social networks (OSNs) provide Internet users with the opportunity to discuss, get informed, express themselves, and interact for a myriads of goals, such as planning events and engaging in commercial transactions. In a word, users rely on online services to say to the world what they are, think, do; and viceversa, they learn the same about the other subscribers.

Quite naturally, the widespread availability and ease of use have made OSNs the ideal setting for the proliferation of fictitious and malicious accounts. While hiding a real identity is sometimes motivated by the harmless side of one's personality, there exist however deceitful situations where accounts of social platforms are created and managed to distribute unsolicited spam, advertise events and products of doubtful legality, sponsor public characters and, ultimately, lead to a bias within the public opinion~\cite{liu2014}. Nonetheless, the plague of such social spammers and bots leads to an ingenious and lucrative ``underground economy", where account vendors, their customers, and oblivious victims play a piece staging since the early 00s~\cite{Stringhini:2012, Stringhini:2013,Thomas2013}. 

Peculiarity of social spambots is that they evolve over time, adopting sophisticated techniques to evade early-established detection approaches, such as those based on textual content of shared messages~\cite{Lee:2010}, posting patterns~\cite{Stringhini:2010}, and social relationships~\cite{fortunato2010, Ghosh:2012}. As evolving spammers became clever in escaping detection, for instance by changing discussion topics and posting activities, researchers stayed in line with the times and proposed complex models based on the interaction graphs of the accounts under investigation~\cite{yang2013, hu2014}.

Noticeably, spambots evolution still goes on. Recent investigations highlight how new waves of social spambots are rising~\cite{ferrara2016}. Their characteristics are such that a standard classification approach, where the single account is evaluated according to a set of established features tested over known datasets, is no longer successful. Instead, the intuition is that the key factor for spotting new social spambots is focusing on the ``collective behavior" of groups of accounts, rather then on single behaviors.

\noindent \textbf{Contributions.} In this work, we contribute along the following dimensions. 

\noindent \textit{Online behavioral modeling:} We propose a strikingly novel, simple and effective
approach to model online users behaviors, targeted to social spambots detection. Behaviors are modeled via \mbox{{\it digital DNA}}, namely strings of characters, each of them encoding one action of the online account under investigation. Similarly to biological DNA, digital DNA allows a compact representation of information. In contrast, the characters encoding the user actions are not restricted to four (as in the case of the four nucleotide bases). Thus, digital DNA is a flexible model, able to represent different actions, on different social platforms, at different levels of granularity.
We extract and analyze digital DNA sequences from the behaviors of OSNs users, and we use
Twitter as a benchmark to validate our proposal. We obtain a
compact and effective DNA-inspired characterization of user
actions. Then, we apply standard DNA analysis techniques to
discriminate between genuine and spambot accounts on Twitter. Throughout the paper, our detection technique -- based on digital DNA modeling of accounts behaviors -- is called \mbox{{\it Social Fingerprinting}}.

\noindent \textit{Spambot detection:} An experimental campaign supports our application. Starting from two Twitter datasets where genuine and spambot accounts are a priori known, we leverage digital DNA  to let recurrent patterns emerge. We show how groups of spambots share common patterns, as opposite to groups of genuine accounts. As a concrete application of this outcome, we demonstrate how to apply our Social Fingerprinting methodology to tell apart spambots from genuine accounts, within an unknown set of accounts. The excellent performances obtained in terms of standard classifiers-based indicators (like F-Measure, Accuracy, Precision, and Recall) support the quality and viability of the Social Fingerprinting technique. 

\noindent \textit{Flexibility and applicability:} While Twitter spambot detection is a specific use case on a
specific social network, our proposed Social Fingerprinting technique is platform and
technology agnostic, hence paving the way for diverse behavioral
characterization tasks.
Indeed, we believe that the high flexibility and applicability of digital DNA sequences make this new modeling approach suitable to represent different scenarios, with the potential to open new directions of research. As example, we cite here
the capability to 
let behavioral patterns emerge from the crowd, as approached with different techniques in~\cite{jiang2016, li2014}. 
Making use of standard DNA sequences alignment tools, 
our approach has the comfortable outcome
of avoiding the often frustrating intervention of humans,
who may not have the means to discriminate patterns by
simply inspecting on an \textit{account by account} basis.

\noindent \textbf{Roadmap.} The remainder of this paper is as follows. Section~\ref{sec:related-works} presents a survey of relevant work in the field of social networks spambot detection. The Twitter datasets of our experiments are introduced in Section~\ref{sec:twitter-dataset}, while, in Section~\ref{sec:digital-dna}, we introduce the notion of digital DNA and we propose a similarity measure for digital DNA sequences.  Section~\ref{sec:char-account} depicts the characteristics of online accounts based on their DNA sequences. Section~\ref{sec:best-cut-choice} presents the analysis and the results of our approach for Twitter spambot detection. In Section~\ref{sec:discussion}, we discuss the experimental results, the lessons learned, and the generality of the proposed approach. Finally, Section~\ref{sec:conclusions} draws the conclusions.

\makeatletter{}
\section{Related work}
\label{sec:related-works}
Ranged over a period spanning  the last six years, the academic literature has seen the flowering of scientific approaches to model and analyze anomalous accounts on social networks. In particular, Twitter has gained a lot of attention, since the platform massively features different kinds of peculiar subscribers, such as {\it spammers}, {\it bots},  {\it cyborgs}, and {\it fake followers}. 

In a nutshell, spammers are those accounts that advertise un-solicited and often harmful content, containing links to malicious pages~\cite{Stringhini:2010}, bots are computer programs
that control social accounts, as stealthy as to mimic real users~\cite{SocialBot11}, while cyborgs interweave
characteristics of both manual and automated behavior~\cite{ChuGWJ:2012}. Finally, there are fake followers, namely
accounts massively created to follow a target account and that can be bought from online markets~\cite{Stringhini:2013,cresci2015}, also attracting the interest of mass media, with sometimes questionable results \cite{DASec:2014}. 
Each of these categories has been the matter of several investigations. 

\subsection{Established techniques}
As an example for spam detection, a branch of research mined the textual content of
tweets~\cite{GaoCLPC12}, others studied the redirection of embedded URLs in tweets~\cite{Lee:2013}, or
classified the URLs landing pages~\cite{ThomasGMPS11}. Work in~\cite{Gao14} moves beyond the difficulty of labeling those
tweets without URLs as spam tweets, by proposing a composite tool, able to match incoming tweets with
underlying templates commonly used by spammers.

Other work investigated spammers  through a multi-feature approach,
including features on the profile, the behavior, and the timeline of an account. Examples of such an analysis include~\cite{Stringhini:2010,yang2013,weibo14}.  In particular, \cite{yang2013} designed a series of novel criteria, demonstrating their efficacy in detecting those spammers that evade existing detection techniques. 

Our previous work in~\cite{cresci2015} considers fake Twitter followers. Since both spammers, bots, and genuine accounts could fall in this category, we tested a series of rules and features from both the grey literature and Academia on a reference dataset of humans and fake followers. Our main contributions were: (i) pruning those rules and features that behaved worst in detecting fake followers, and (ii) implementing a classifier which significantly reduces overfitting and cost for data gathering. 

The above listed contributions are an excerpt of the research efforts towards malicious accounts detection over the last recent years. 
Noticeably, the majority of those contributions was grounded on the assumption that it is possible to recognize an account as genuine or not, based on a series of characteristics featured by that account. The classification takes place on the single account, and the classification results hold since ranges of values for such characteristics have been previously proved on reference datasets to be symptomatic of anomalies. 

\subsection{Emerging trends}
Remarkably, we observe a significant shift taking place over the last two years. As observed in~\cite{ferrara2016}, new {\it social bots} are rising, whose peculiarity emerges only when considering  their collective behavior.  As observed from the analysis of the datasets introduced in Section~\ref{sec:twitter-dataset}, the new waves of social bots are such that, if the accounts are considered one by one, they are no more distinguishable from genuine ones. We claim that such social spambots represent the third and most novel generation of spambots, following the original wave dating back from the 00s, passing through a second generation dated around 2011 (described by Yang {\it et al.} in~\cite{yang2013}). Interestingly, the thesis of a third, recent generation of spambots is also supported by related work carried on over datasets dated around the early 10s, such as~\cite{yang2014uncovering}, whose conclusions were that malicious accounts in a group appeared as disconnected among themselves and whose behaviors were not similar between one another.

In this work, we show how digital DNA represents a powerful basis for the detection of the third generation of spambots. Noticeably, we acknowledge the publication of a few recent work whose background philosophy is aligned with ours. Indeed, such works consider different shades of behavioral characteristics of the accounts, with the commonality to study them  as a group instead of one by one. In the following, we relate on such novel work, highlighting differences and similarities with ours.

Work in~\cite{Beutel:2013} and~\cite{jiang2015} study connectivity patterns
in large graphs to let unexpected behaviors emerge. Following the factual data that unexpected behaviors feature lockstep characteristics, e.g., large groups of followers connect to the same groups of followees, the authors depict correspondences between lockstep behaviors in the social graph and  dense blocks in the adjacency matrix of the graph. Furthermore, they propose  an algorithm to spot users featuring unexpected behaviors.

The authors of~\cite{Giatsoglou2015} move from the intuition that, if a collective online action happens once, then that action is not necessarily fraudulent. 
Instead, if that collective action repeats over time, especially in reaction to the same kind of event, it probably represents an anomalous activity. In particular, the work focuses on retweeting activities, defines features for retweet threads characterization, and proposes a methodology for catching synchronized frauds.

SynchroTrap~\cite{Cao:2014} aims at detecting loosely synchronized behaviors for a broad
range of social network applications. Time is an important dimension for SynchroTrap (thus marking a difference with the approach proposed here), since the methodology forms clusters on the basis of equal actions performed by online accounts within the same time interval.

Inspired by particle physics, fluids mechanics, and astronomy, the authors of~\cite{yu2015}  consider group anomalies in a wider flavor, not necessarily oriented to groups of spambots. As an example, focusing on the major conferences in the area of artificial intelligence, they consider if there are published papers whose topics are anomalous for those conferences, leveraging features inherent to both the single component (e.g., the topic of the paper) and the relationships among components (e.g., authors common to different papers).

\subsection{Comparison}
We briefly  highlight the main differences of our approach with respect to the cited papers. First of all, we consider a single dimension as the basis to let groups of social accounts emerge: the digital DNA, i.e., the sequence of characters encoding the accounts' behavior. Secondly, in this work we do not consider properties of the social graph (e.g., a follow link over Twitter or a friendship on Facebook). This leads to the significant advantage of reducing the cost for data gathering. Indeed, approaches that are based on graph mining (such as~\cite{jiang2016}) generally rely on a large quantity of data and can require computationally expensive algorithms to perform their detection~\cite{cresci2015}. Our proposal, instead, only exploits Twitter timeline data to perform spambots detection.

Third, we enable analysts to leverage a
powerful set of tools -- developed over decades
for DNA analysis -- to validate their working hypotheses on
online spambots behaviors.

Furthermore, our DNA-inspired modeling focuses on the concept of sequence, namely ordered lists of symbols, with variable length, 
and taken from a relatively small alphabet. This marks a clear separation from other well-known
behavioral analysis techniques that do not consider the ordering of the elements, like hashing~\cite{ou2015}.

In Section~\ref{sec:discussion}, we will show a comparison of our approach with two unsupervised approaches, namely \cite{ahmed2013} and \cite{miller2014}, in terms of detection performances. As discussed later on, the results are promising and they lead us to believe that digital DNA is a simple and compact, yet powerful, mean to detect the novel waves of social spambots. 
The intuition behind our approach has been succinctly presented in a magazine paper~\cite{IntSys2015}.

\makeatletter{}
\section{Twitter datasets}
\label{sec:twitter-dataset}
In this section, we describe the different Twitter
datasets that constitute the real-world data used in our
experiments.
Specifically, we collected some months' worth of data
about the activities of a random sample of genuine (human-operated) accounts
and of two different families of spambots.
\begin{figure}[!tb]
  \centering
  \subfigure[Twitter profile of a social spambot, belonging to the \texttt{Bot1} group.
  \label{fig:account_spambot1}]{\includegraphics[width=0.24\textwidth]{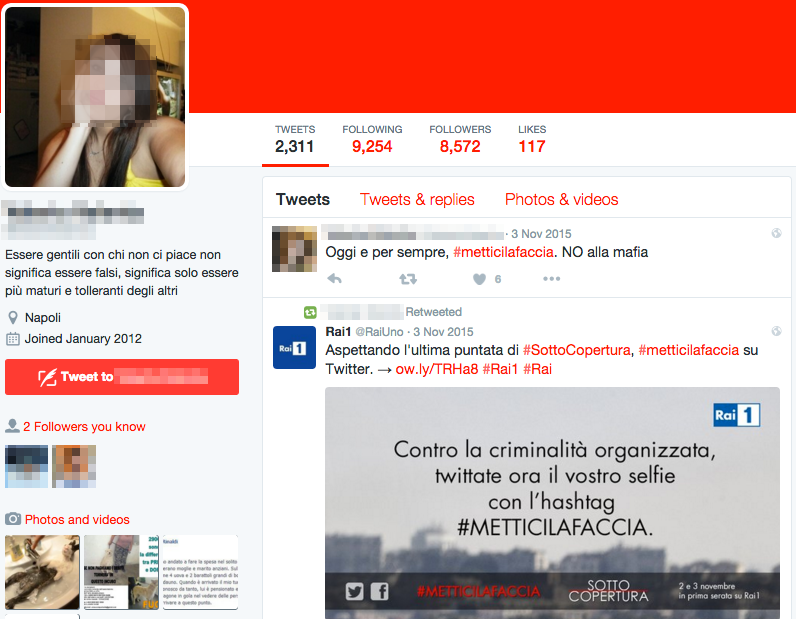}}
  \subfigure[Twitter profile of a social spambot, belonging to the \texttt{Bot2} group.
  \label{fig:account_spambot2}]{\includegraphics[width=0.24\textwidth]{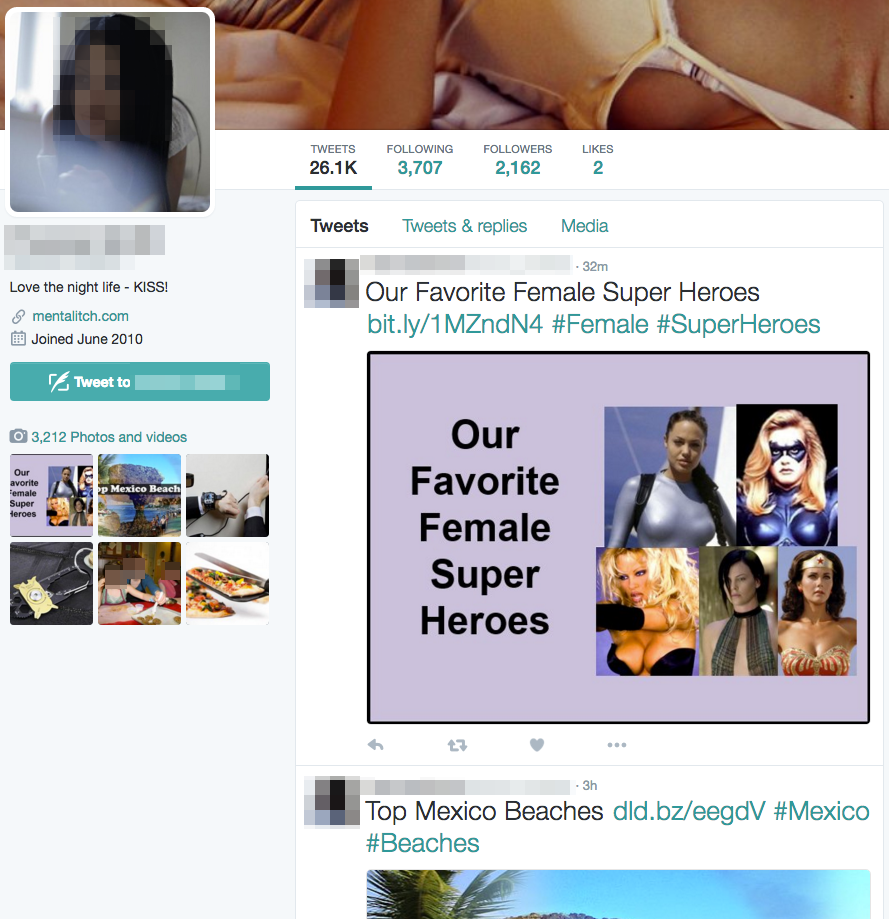}}\\
  \subfigure[Twitter profile of a genuine user.
  \label{fig:account_genuine}]{\includegraphics[width=0.24\textwidth]{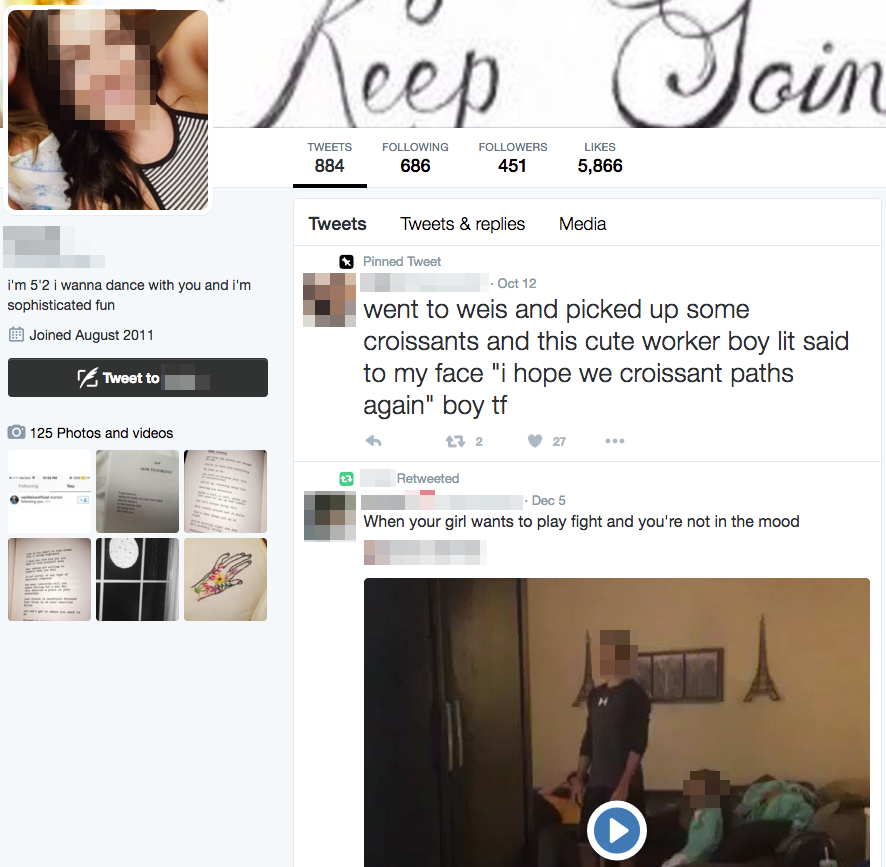}}
  \caption{Examples of Twitter profiles of social spambots and genuine users. The carefully engineered profiles of the novel spambots make them nearly indistinguishable from genuine users.
  \label{fig:account_screenshots}}
\end{figure}
\makeatletter{}
\begin{table*}[!tb]
	\footnotesize
	\centering
	\begin{tabular}{lrrrrrr}
		\toprule
		&&& \multicolumn{3}{c}{{\textbf{relationships}}} &\\
		\cmidrule{4-6}
		\textbf{dataset} & \textbf{accounts} & \textbf{tweets} & followers & friends & total & \textbf{interactions} \\
		\midrule
		\texttt{Bot1} (retweeters of the political candidate)		& 991	& 1,610,176	& 4,031,897	& 4,022,884	& 8,054,781	& 147,387 \\
		\texttt{Bot2} (spammers of \textit{Amazon.com} products)	& 464	& 1,418,626	& 1,352,370	& 818,913		& 2,171,283	& 15,041 \\
		human										& 3,474	& 8,377,522	& 4,740,286	& 2,153,107	& 6,893,393	& 591,768 \\
		\bottomrule
	\end{tabular}
	\caption{Statistics about the Twitter datasets.
	\label{tab:datasets}}\vspace{-0.5cm}
\end{table*}

A first dataset of spambots was created after observing the activities of a novel group of social bots that we discovered on Twitter during the last Mayoral election in Rome, in 2014. One of the runners-up employed a social media marketing firm for his electoral campaign, that made use of almost 1,000 automated accounts on Twitter to publicize his policies. Surprisingly, we found such automated accounts to be similar to genuine ones in every way. Every profile was accurately filled with detailed -- yet fake -- personal information such as a stolen photo, short-bio, location, etc. Those accounts also represented credible sources of information since they all had thousands of followers and friends, the majority of which were genuine users. Furthermore, the accounts showed a tweeting behavior which was apparently similar to those of genuine accounts, with a few tweets posted every day, mainly quotes from popular people. However, every time the political candidate posted a new tweet from his official account, all the automated accounts retweeted it in a time span of just a few minutes. By resorting to this farm of bot accounts, the political candidate was able to reach many more genuine accounts in addition to his direct followers and managed to alter Twitter engagement metrics during the electoral campaign.
Amazingly, we also found tens of human accounts who tried to engage in conversation with some of the spambots. The most common form of such human-to-spambot interaction was represented by a human reply to one of the spambot tweets quotes. Quite obviously, no human account who tried interacting with the spambots ever received a reply from them.

We further investigated this issue and found it to be widespread also outside Italy.
Indeed, we uncovered a second group of social bots whose intent was to advertise a subset of products on sale on the \textit{Amazon.com} e-commerce platform. This time the deceitful activity was carried out by spamming URLs pointing to the advertised products. However, similarly to the retweeters of the Italian political candidate, also this family of spambots interleaved spam tweets with many harmless and genuine ones. 

Henceforth, we refer to the spambots retweeters of the Italian political candidate as \verb|Bot1| and to those spambots advertising \textit{Amazon.com} products as \verb|Bot2|. In order to give more insights into the advanced characteristics of the novel social spambots, in Figure~\ref{fig:account_screenshots} we show the profile pages of 2 spambots belonging to the \texttt{Bot1} and the \texttt{Bot2} groups and that of a genuine account. As shown, from a mere comparison of the Twitter profiles, it is nearly impossible to tell apart the spambots from the genuine account. Worryingly, this is the same scenario that Twitter users are typically presented to, while browsing the social platform. To make the situation even worse, Figures~\ref{fig:account_spambot1} and~\ref{fig:account_spambot2} show that the novel social spambots also employ social engineering techniques, such as the profile picture of a young attractive woman and the occasional posting of provocative tweets, in order to lure genuine accounts. As such, any threat spread out by social spambots (e.g., malware, phishing attacks, etc.) is more likely to result in a successful attack with respect to those spread by traditional spambots.

After identifying possible spambots, we exploited a Twitter crawler to
collect data about all the accounts we suspected to belong to the two
groups of spambots. All the accounts collected in this process have
then undergone a manual verification phase to certify their automated
nature. Specifically, the spambots of our datasets were annotated by two tech-savvy post-graduate students, with yearly experience on Twitter and social media.
To evaluate the inter-annotator agreement, we used the well-known Cohen's Kappa ($\kappa$) evaluation metric~\cite{gwet2014}.
For the accounts of the \texttt{Bot1} group, $\kappa = 0.824$, while for the accounts of the \texttt{Bot2} group, $\kappa = 0.351$.
The two values are considered, respectively, excellent and fair~\cite{landis1977measurement}. The disagreements between the two annotators have been resolved by a super-annotator, i.e., a Ph.D. student with yearly experience in sybil and spambot detection.
Summarizing, among all the distinct retweeters of the Italian political
candidate, 50.05\% (991 accounts) were certified as
spambots. Similarly, 89.29\% (464 accounts) of the accounts that
tweeted suspicious \textit{Amazon.com} URLs were also certified as
spambots.
These 2 sets of accounts represent our ground truth of social
spambots. 

Then, in order to build a dataset of certified human
accounts, we randomly contacted Twitter users by asking them simple questions in natural
language, following a hybrid crowdsensing approach~\cite{avvenuti2017}.
Replies to our questions were manually verified and all the
3,474 accounts that answered were certified as humans. For all the
4,929 accounts of our datasets, we then collected behavioral data by
crawling the content of their Twitter pages. Furthermore, we also
collected data about all their direct followers and friends, and about
all the accounts they interacted with in their
tweets. Table~\ref{tab:datasets} shows some statistics about total
collected data.

\makeatletter{}
\section{Digital DNA}
\label{sec:digital-dna}
The human genome is the complete set of genetic information on humans and it is encoded in the form of nucleic acid (DNA) sequences.
A DNA sequence is a succession of characters (i.e., a string) that indicates the order of nucleotides within a DNA molecule. The possible characters are \texttt{A}, \texttt{C}, \texttt{G}, and \texttt{T}, representing the four nucleotide bases of a DNA strand: adenine, cytosine, guanine, thymine. Biological DNA stores the information which directs functions and characteristics of a living organism.
Nowadays, DNA sequences are exploited worldwide in biomedical science, anthropology, forensics, and other branches of science.
DNA sequences can be read from raw biological material through DNA sequencing methods. Currently, such sequences are stored in sequence databases and they are analyzed by means of bioinformatics techniques. Among the most well-known and widely adopted analysis techniques are sequence alignment and repetition/motif elicitation. One of the main goals of these techniques is to find commonalities and repetitions among DNA sequences. Indeed, via an analysis of common sub-sequences and substrings it is possible to predict specific characteristics of the individual and to uncover relationships between different individuals.

By drawing a parallel with biological DNA, we envisage the possibility to model OSNs users behaviors and interactions by means of strings of characters, representing the sequence of their actions. Indeed, online actions -- such as posting new content, replying to another user, following an account -- can be encoded with different characters, similarly to DNA sequences, where the \texttt{A}, \texttt{C}, \texttt{G}, \texttt{T} characters encode the four nucleotide bases. According to this parallelism, a user's actions represent the bases of his/her digital DNA.
As highlighted in~\cite{zafarani2014}, there exist different kinds of user behaviors on OSNs. Digital DNA is a flexible and compact -- yet effective -- way of modeling such behaviors. Its flexibility lies in the possibility to choose which actions to consider while building the DNA sequence. For example, a digital DNA sequence can be built to model user-to-user interactions on Facebook by defining a different base for every possible interaction type, such as comments (base \texttt{C}), likes (base \texttt{L}), shares (base \texttt{S}) and mentions (base \texttt{M}). Users interactions can then be encoded as strings composed of the \texttt{C}, \texttt{L}, \texttt{S} and \texttt{M} characters according to the sequence of actions they perform.
Similarly, it is possible to model users tweeting behaviors on Twitter by defining different bases for tweets, retweets, and replies. Users tweeting behaviors can then be encoded as a sequence of characters according to the sequence of tweets they post.
To this regard, digital DNA shows a major difference with biological DNA where the four nucleotide bases are fixed. In digital DNA both the number and the meaning of the bases can change according to the behavior/interaction one aims to model. Similarly to its biological counterpart, digital DNA is also a compact representation of information. For example, the timeline of a Twitter user can be encoded as a single string of 3,200 characters (one character per tweet).

There is a vast number of algorithms and techniques to draw upon for the analysis of digital DNA sequences. Indeed, many of the techniques developed in the last few years in the field of bioinformatics for the analysis of biological DNA can be leveraged to study the characteristics of digital DNA as well.

In the following we give a general definition of digital DNA, and we introduce the Twitter's digital DNA concept, with one of its possible applications: spambots detection.

\subsection{Definition of digital DNA sequences and its application to Twitter}
\label{sec:twitter-dna}
\newcommand{\Bttype}{\ensuremath{\mathbb{B}^3_{\textit{type}}}}
\newcommand{\Btcontent}{\ensuremath{\mathbb{B}^3_{\textit{content}}}}
\newcommand{\Bscontent}{\ensuremath{\mathbb{B}^6_{\textit{content}}}}

The bases used to create a digital DNA sequence are represented as a finite set of unique symbols or characters, denoted by $\mathbb{B}$ and defined as:
\begin{equation*}
\label{eq:alphabet-def}
\mathbb{B} = \{B_1, B_2, \ldots, B_N\} \quad B_i \neq B_j \;\; \forall \;\; i,j = 1, \ldots, N \; \wedge \; i \neq j
\end{equation*}
The set $\mathbb{B}$ is also called the \textit{alphabet} of a digital DNA sequence. The number of bases used to create a sequence is the cardinality of its alphabet, $N = |\mathbb{B}|$.
A digital DNA sequence is an ordered tuple, or row vector, of characters (i.e., a string) whose possible values are defined by the bases of its alphabet. A sequence $s$  is defined as:
\begin{equation*}
\label{eq:sequence-def}
s = (b_1, b_2, \ldots, b_n) \quad b_i \in \mathbb{B} \;\; \forall \;\; i = 1, \ldots, n
\end{equation*}
The number of actions encoded in a DNA sequence determines the length of the sequence, $n = |s|$. Thus, a limited number of bases in an alphabet can be used to create sequences of arbitrary length.

Encoding someone's behavior in a digital DNA sequence means linking each of the actions one aims to model, to a base of the alphabet.
For instance, one can scan someone's actions in chronological order and  assign the appropriate base to each of the actions: the succession of bases generated in this way makes up the digital DNA sequence.
As a practical example, we can model Twitter accounts behavior by defining the following alphabet, of cardinality $N = 3$, based on the \textit{type} of tweets produced:
\begin{equation*}
\label{eq:tweettype3}
\Bttype = 
\left \{
\begin{aligned}
&\verb|A| \mapsfrom \text{\scalebox{0.7}{tweet}}, \\
&\verb|C| \mapsfrom \text{\scalebox{0.7}{reply}}, \\
&\verb|T| \mapsfrom \text{\scalebox{0.7}{retweet}}
\end{aligned}
\right \} = \{\verb|A|, \verb|C|, \verb|T| \}
\end{equation*}
A digital DNA sequence based on the \Bttype\ alphabet can then be obtained by scanning the tweets produced by a user on Twitter and by assigning the \verb|T| character to every retweet, the \verb|C| character to every reply, and the \verb|A| character to every other tweet, in the same order of the tweets generated by the user.
An excerpt of a digital DNA sequence generated with the alphabet \Bttype\ is $s = (\verb|A|,\verb|A|,\verb|A|,\verb|C|,\verb|A|,\verb|T|,\verb|C|,\verb|A|,\verb|A|,\verb|C|, \ldots)$. A digital DNA sequence can also be represented with a more compact notation as a string $s = \verb|AAACATCAAC|\ldots \;\;$, instead of a row vector.

The \Bttype\ alphabet represents just one among the possible ways of modeling users behaviors on Twitter. Other ways of modeling such behaviors can draw upon the content of tweets, rather than their type. 
For instance, in order to classify a tweet based on its content we exploited Twitter's notion of entities\footnote{Twitter defines the following types of entities: URLs, \#hashtags, @mentions and media (images, videos). For a complete reference of Twitter entities, see: \url{https://dev.twitter.com/overview/api/entities}}. We exploited Twitter entities to define the \Btcontent\ and \Bscontent\ alphabets, which can be used to model the content of tweets with different degrees of granularity, as in the following:
\begin{equation*}
\label{eq:tweetcontent3}
\Btcontent =
\left \{
\begin{aligned}
&\verb|N| \mapsfrom \text{\scalebox{0.7}{tweet contains no entities (plain text)}}, \\
&\verb|E| \mapsfrom \text{\scalebox{0.7}{tweet contains entities of one type}}, \\
&\verb|X| \mapsfrom \text{\scalebox{0.7}{tweet contains entities of mixed types}}
\end{aligned}
\right \} = \{\verb|N|, \verb|E|, \verb|X| \}
\end{equation*}
\label{eq:tweetcontent6}
\begin{align*}
\ensuremath{\Bscontent}  & \ensuremath{=\left \{
    \begin{aligned}
          \texttt{N} &\mapsfrom \text{\scalebox{0.7}{tweet contains no entities (plain text)}}, \\
          \texttt{U} &\mapsfrom \text{\scalebox{0.7}{tweet contains one or more URLs}}, \\
          \texttt{H} &\mapsfrom \text{\scalebox{0.7}{tweet contains one or more hashtags}}, \\
          \texttt{M} &\mapsfrom \text{\scalebox{0.7}{tweet contains one or more mentions}}, \\
          \texttt{D} &\mapsfrom \text{\scalebox{0.7}{tweet contains one or more medias}}, \\
          \texttt{X} &\mapsfrom \text{\scalebox{0.7}{tweet contains entities of mixed types}}
      \end{aligned}
      \right\}= \phantom{MMM}}\\ 
      &\ensuremath{= \{\texttt{N}, \texttt{U}, \texttt{H}, \texttt{M}, \texttt{D}, \texttt{X} \} }
\end{align*}

Another possible way of modeling the content of tweets could have involved the detection of the topic of a tweet~\cite{sriram2010, lee2011}. Then, it would have been possible to define an alphabet so as to have a different base for each of the main topics, such as politics, sports, technology, music, etc. Anyway, for the sake of simplicity, in our work we only exploited Twitter entities in order to obtain DNA sequences based on the content of tweets.

In the above notations, alphabets are characterized by a subscript (e.g., $\textit{type}$) that identifies the kind of information captured by the bases, and by a superscript (e.g., $3$) that denotes the number $N$ of bases in the alphabet. These two indices are typically enough to unequivocally identify an alphabet. As demonstrated by the \Btcontent\ and \Bscontent\ alphabets, the superscript is useful to distinguish alphabets modeling the same facet with a different number of bases.

\subsection{LCS: a similarity measure for digital DNA sequences}
\label{sec:simmeas}

As seen above, a digital DNA sequence is a data representation that is
suitable to model the behavior of a single OSN user.  However, when
analyses are targeted to groups rather than single users, it could be
useful to manage and study multiple digital DNA sequences as a whole, in
order to infer the characteristics of the group.  Here, we study
collective behaviors via an analysis of the similarities among the
digital DNA sequences of the users of a given group.  A group $A$ of
$M = | A |$ users can be described by the digital DNA sequences of the
$M$ users, namely:
\begin{equation*}
\label{eq:accounts}
A = \colvect[c]{s_1;s_2;\vdots;s_M} = \colvect[c]{(b_{1,1}, b_{1,2}, \ldots, b_{1,n});(b_{2,1}, b_{2,2}, \ldots, b_{2,m});\vdots;(b_{M,1}, b_{M,2}, \ldots, b_{M,p})}
\end{equation*}
In the above characterization, the group $A$ is defined as a column
vector of $M$ digital DNA sequences of variable length, one sequence
for each user of the group.

Many algorithms and techniques have been developed in recent years for
the analysis of biological DNA sequences or, more generally,
strings. Such techniques mainly come from the fields of bioinformatics
and string mining~\cite{gusfield1997}. Thus, the adoption of a
behavioral data representation that is based on DNA strings opens up
the possibility to leverage recent advances is such
fields. Furthermore, decades of research and development led to
scalable and efficient algorithms, that fit well with the need to
manage and study OSNs data, which is by its nature,
humongous and ever-growing.
\begin{figure}
  \centering
  \subfigure[\Bttype\ alphabet.
  \label{fig:lcs-tweettype3-human}]{\includegraphics[width=0.24\textwidth]{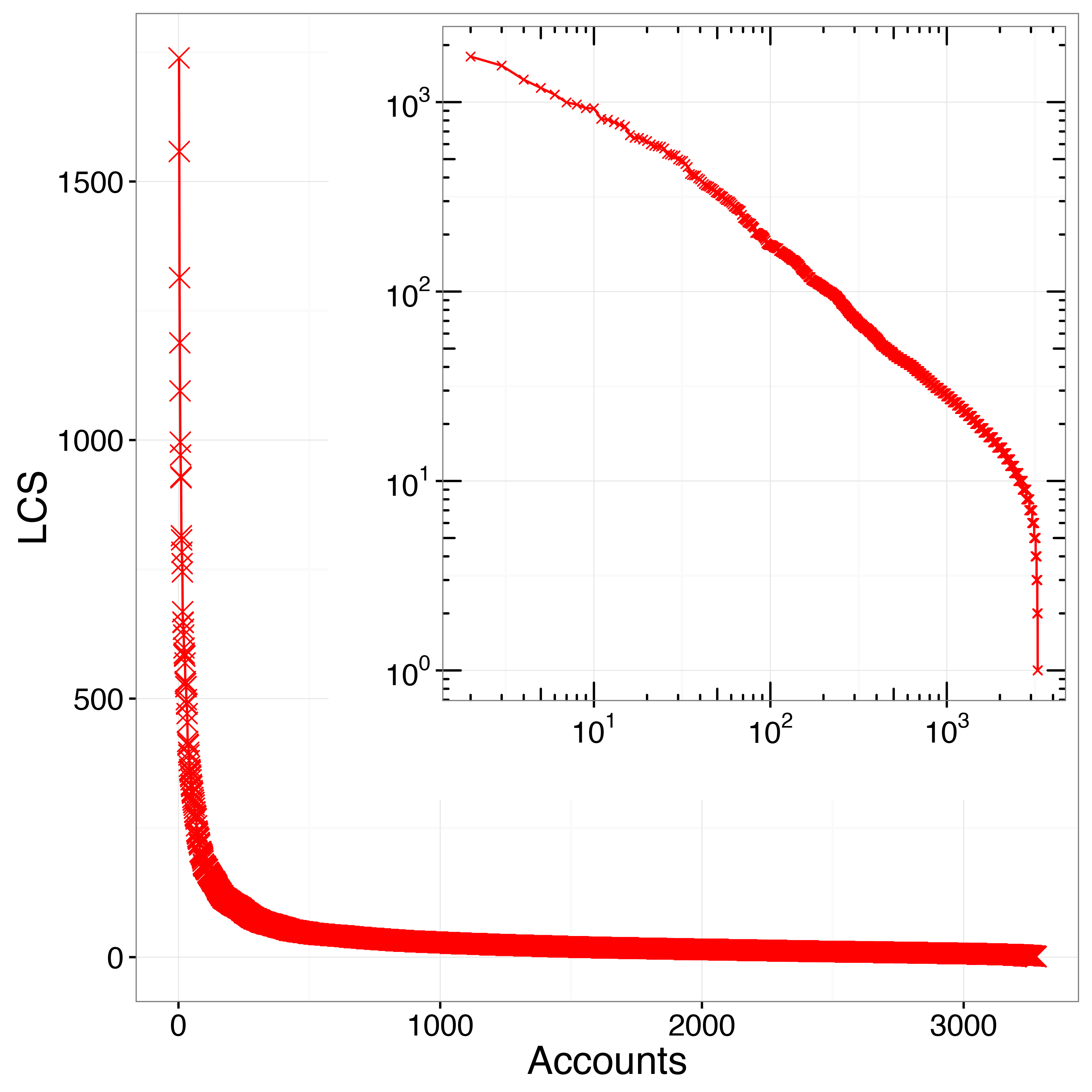}}
  \subfigure[\Btcontent\ alphabet.
  \label{fig:lcs-tweetcontent3-human}]{\includegraphics[width=0.24\textwidth]{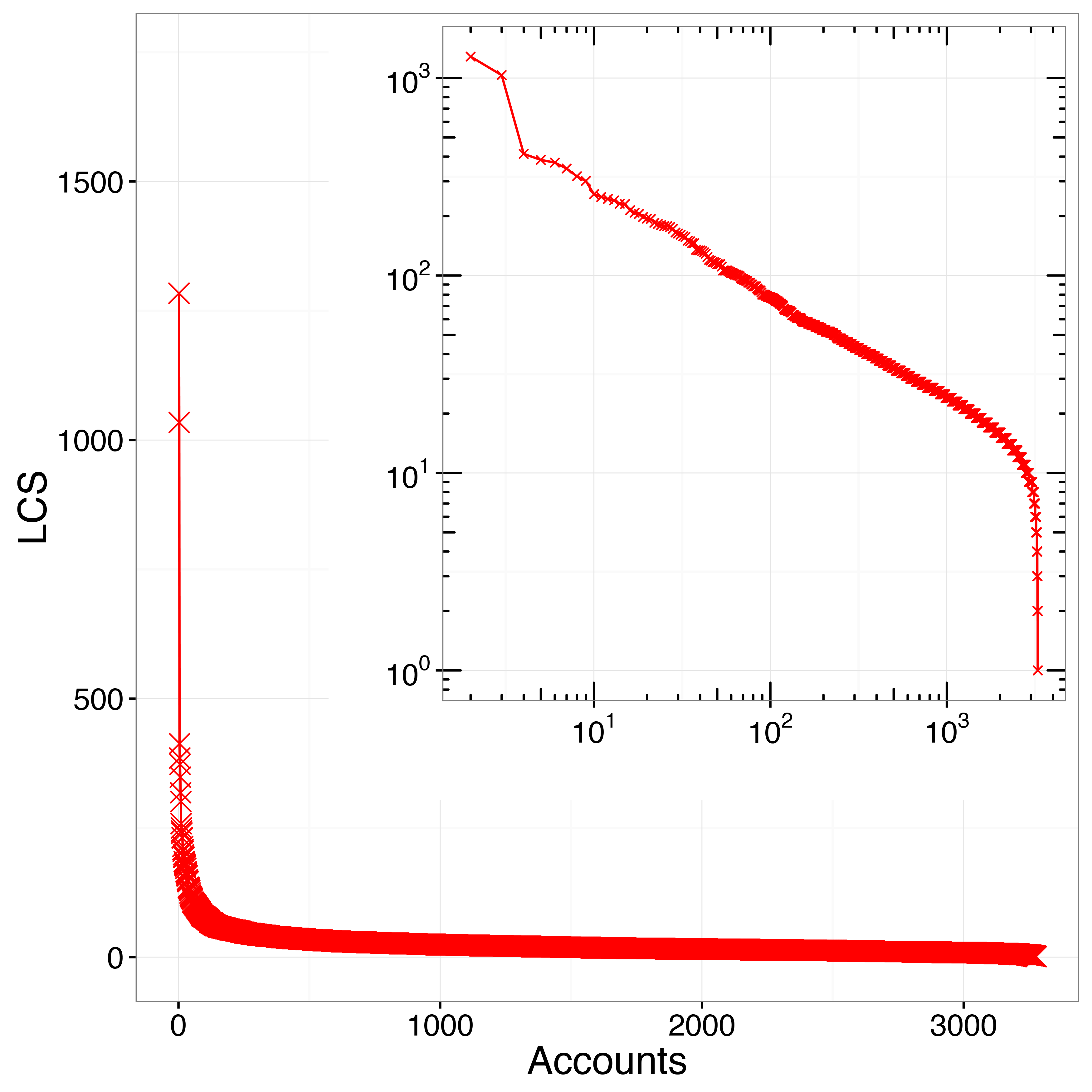}}\\
  \subfigure[\Bscontent ~alphabet.
  \label{fig:lcs-tweetcontent6-human}]{\includegraphics[width=0.24\textwidth]{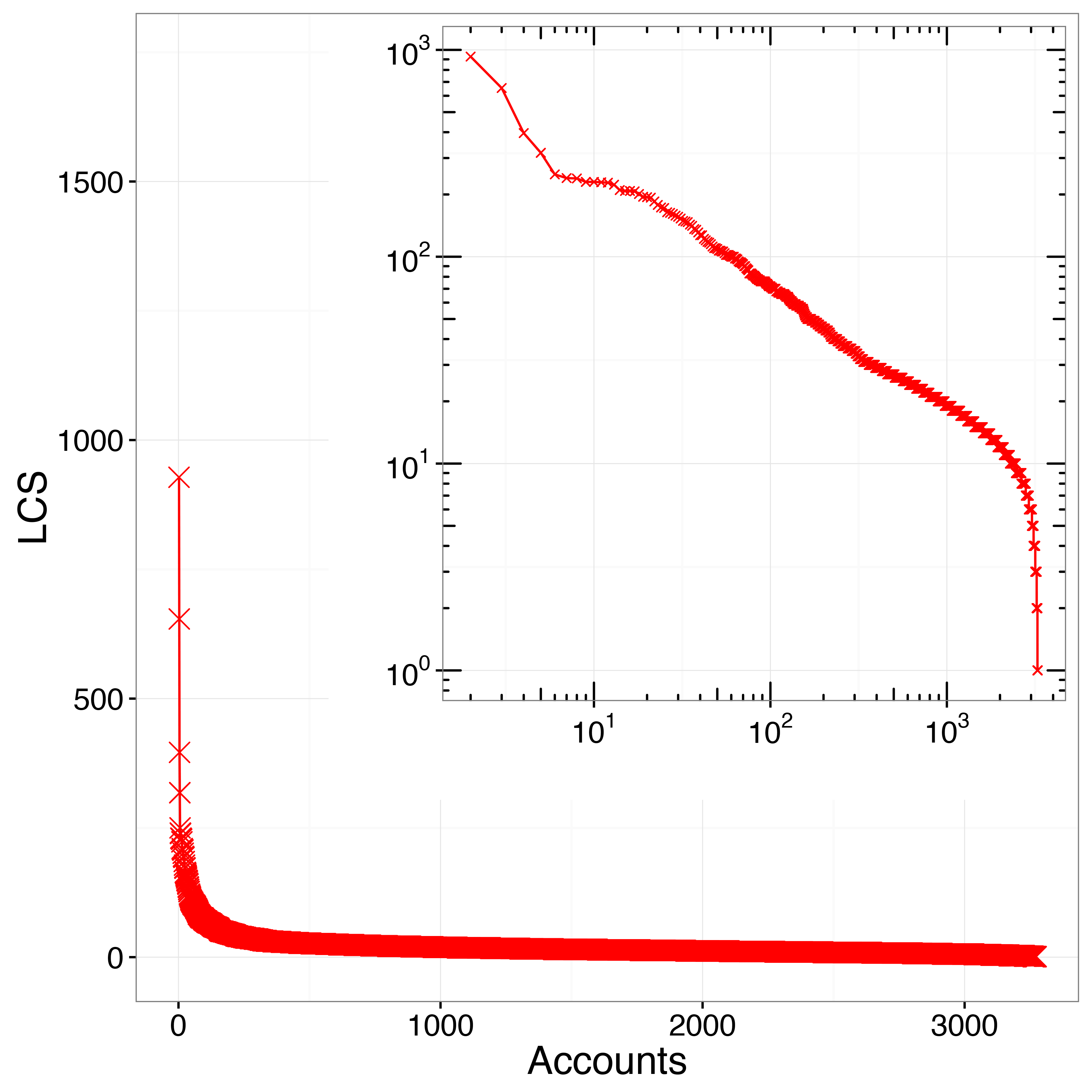}}
  \caption{LCS curves of a group of genuine (human-operated) accounts.
  \label{fig:lcs-human}}
\end{figure}

Among the possible means to quantify similarities between sequential
data representations, in our work we relied on the notion of the
\textit{longest common substring} between two or more DNA
sequences~\cite{arnold2011}.  Intuitively, users that share long
behavioral patterns are much more likely to be similar than those that
share little to no behavioral patterns.  Given two strings, $s_i$ of
length $n$ and $s_j$ of length $m$, their longest common substring
(henceforth LCS) is the longest string that is a substring of both
$s_i$ and $s_j$. For example, given $s_i = \verb|WASHINGTON|$ and $s_j
= \verb|RINGTONE|$, the LCS between $s_i$ and $s_j$ is the string
\verb|INGTON|. The extended version of this problem that considers an
arbitrary finite number of strings, is called the \textit{k-common
  substring} problem~\cite{chi1992}. In this case, given a vector $A =
(s_1, \dots, s_M)$ of $M$ strings, the goal is to find the
LCS that is common to at least $k$ of these strings, for each $2 \le k
\le M$.  Notably, both the \textit{longest common substring} and the
\textit{k-common substring} problems can be solved in linear time and
space, by resorting to the generalized suffix tree and by implementing
state-of-the-art algorithms, such as those proposed
in~\cite{arnold2011}. Solving the LCS \textit{k-common substring} problem for each $2 \le k \le M$, it is possible
to plot an \textit{LCS curve}, showing the relationship between the
length of the LCS and the number $k$ of strings. For example,
Figures~\ref{fig:lcs-tweettype3-human},~\ref{fig:lcs-tweetcontent3-human},
and~\ref{fig:lcs-tweetcontent6-human} depict the LCS curves
computed for a set of genuine (human-operated) Twitter accounts. On the $x$
axis is reported the number of $k$ accounts (corresponding to the $k$
strings, or digital DNA sequences, used to compute LCS values) and on
the $y$ axis the length of the LCS common to at least $k$
accounts. Therefore, every point in an LCS curve corresponds to a
subset of $k$ accounts that share the longest substring (of length
$y$) among all those shared between all the other possible subsets of
$k$ accounts.

As a direct consequence of the definition of LCS, as the number $k$ of
accounts grows, the length of the LCS common to all of them
shortens. In other words, LCS curves are \textit{monotonic nonincreasing} functions:
\begin{equation*}
\text{LCS}[k-1] \ge \text{LCS}[k] \quad \forall \;\; 3 \le k \le M
\end{equation*}
This is also clearly visible in the LCS curves of
Figures~\ref{fig:lcs-tweettype3-human},~\ref{fig:lcs-tweetcontent3-human},
and~\ref{fig:lcs-tweetcontent6-human}. Thus, it is more likely to
find a long LCS among a few accounts rather than among large groups.

\makeatletter{}\section{Characterization of account DNAs with the LCS curves}
\label{sec:char-account}
\begin{figure}
  \centering
  \subfigure[\texttt{Bot1} versus human accounts.
  \label{fig:lcs-hum-vs-bot1}]{\includegraphics[width=0.24\textwidth]{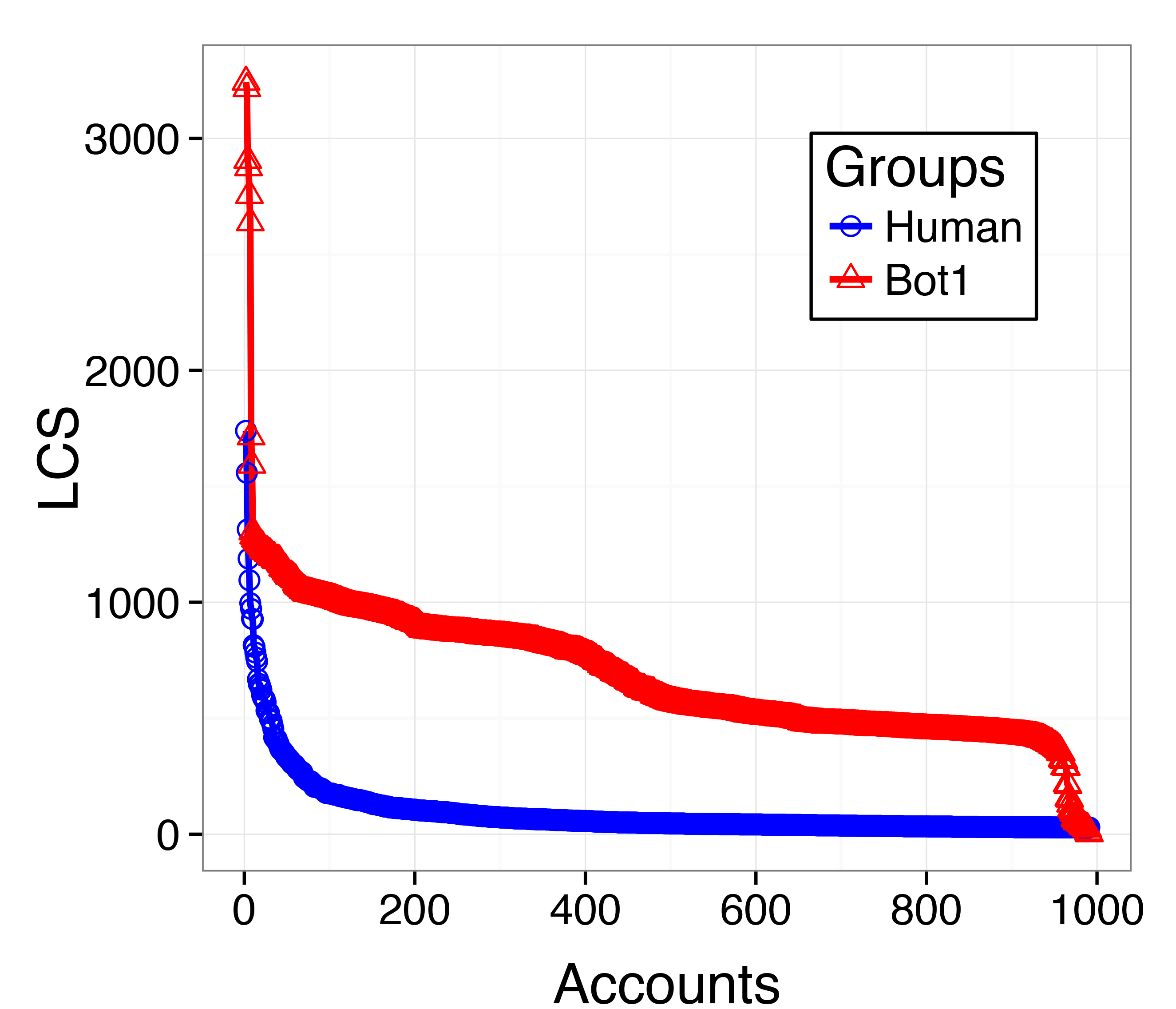}}
  \subfigure[\texttt{Bot2} versus human accounts.
  \label{fig:lcs-hum-vs-bot3}]{\includegraphics[width=0.24\textwidth]{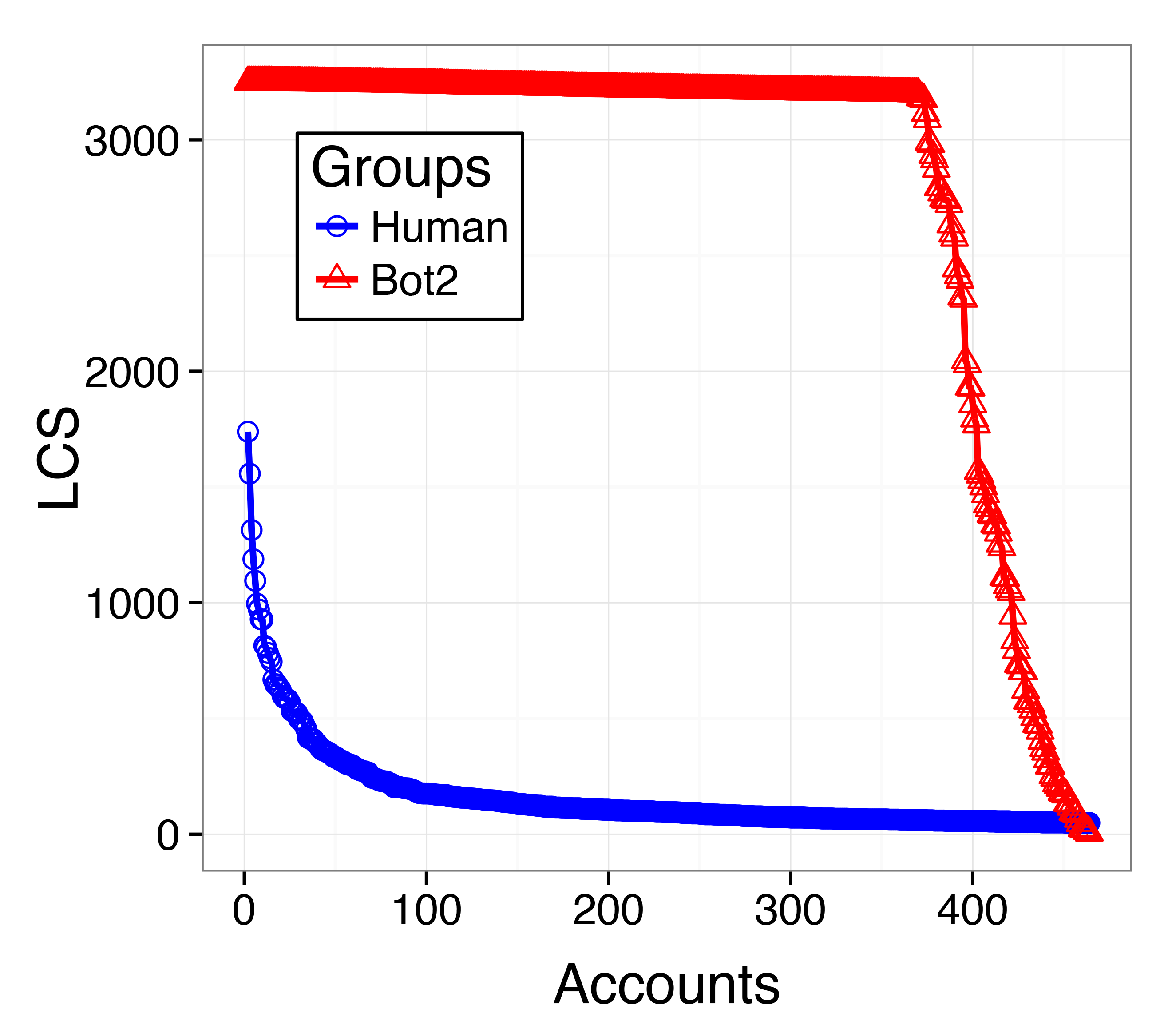}}
  \caption{Comparison between LCS curves of spambots and genuine accounts for the \Bttype\ alphabet.
  \label{fig:lcs-human-vs-bot}}
\end{figure}
To exploit at its best the potential of digital DNA, we need a deeper understanding of the elements that mark the distinction
between genuine users and social spambots. Hence, building on the
definitions of digital DNA and LCS curves given in
Section~\ref{sec:digital-dna}, in this section we study the
characteristics of the LCS curves of the different datasets 
introduced in Section~\ref{sec:twitter-dataset}.  We evaluate the differences and similarities among
those groups of accounts, as seen through the lenses of our digital
DNA sequences.

Figure~\ref{fig:lcs-human-vs-bot} shows a comparison between the LCS
curves of genuine (human) accounts and those of the \texttt{Bot1}
(Figure~\ref{fig:lcs-hum-vs-bot1}) and \texttt{Bot2}
(Figure~\ref{fig:lcs-hum-vs-bot3}) groups.  As shown, the LCS of both
groups of spambots are rather long even when the number of accounts
grows. This is strikingly evident in Figure~\ref{fig:lcs-hum-vs-bot3}
(\texttt{Bot2} -- spammers of \textit{Amazon.com} products). For both
the spambot groups, we observe a sudden drop in LCS length when the
number of accounts gets close to the group size, namely at the end of
the $x$ axis. In contrast to the remarkably high LCS curves of
spambots, genuine accounts show little to no similarity -- as represented
by LCS curves that exponentially decay, rapidly reaching the smallest
values of LCS length.

This preliminary yet considerable differences between the LCS curves of
genuine accounts and spambots suggest that, despite the advanced
characteristics of these novel spambots, the \Bttype\ digital DNA is
able to uncover traces of their automated and synchronized
activity. In turn, the automated behaviors of a large group of
accounts results in exceptionally high LCS curves for such
accounts. Indeed, we consider high behavioral similarity as a proxy
for automation and, thus, an exceptionally high level of similarity
among a large group of accounts might serve as a red flag for
anomalous behaviors. In the following, we preliminarily compare groups
of heterogeneous users, looking for features that could be used to design a
detection mechanism, while in the next section we detail how to leverage
such elements for an effective detection mechanism.

\subsection{LCS curves of a group of heterogeneous users}
\label{sec:heterogeneous-lcs}
\begin{figure}
  \centering
  \subfigure[\texttt{Mixed1} group.
  \label{fig:lcs-mixed1}]{\includegraphics[width=0.23\textwidth]{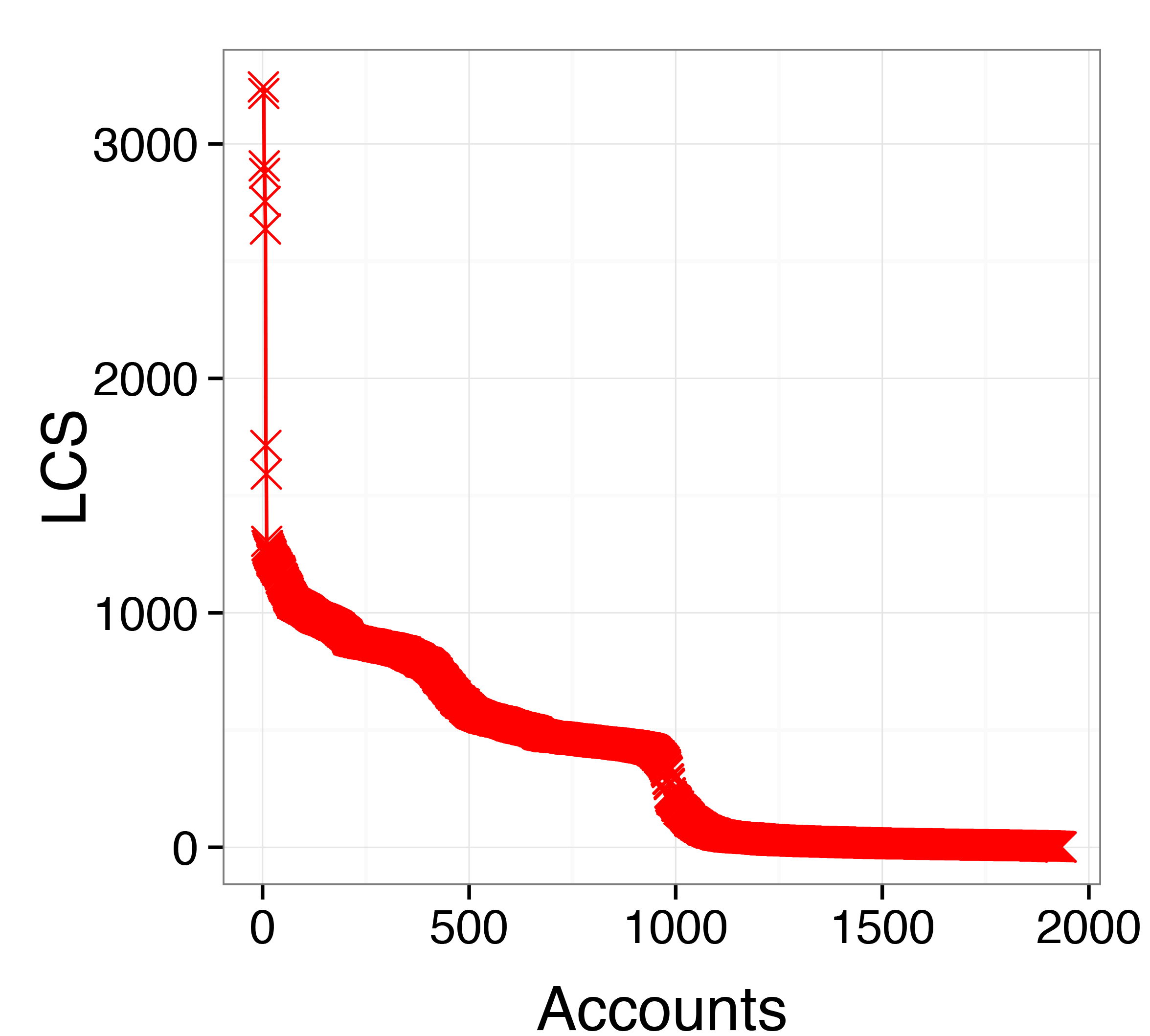}}
  \subfigure[\texttt{Mixed2} group.
  \label{fig:lcs-mixed3}]{\includegraphics[width=0.23\textwidth]{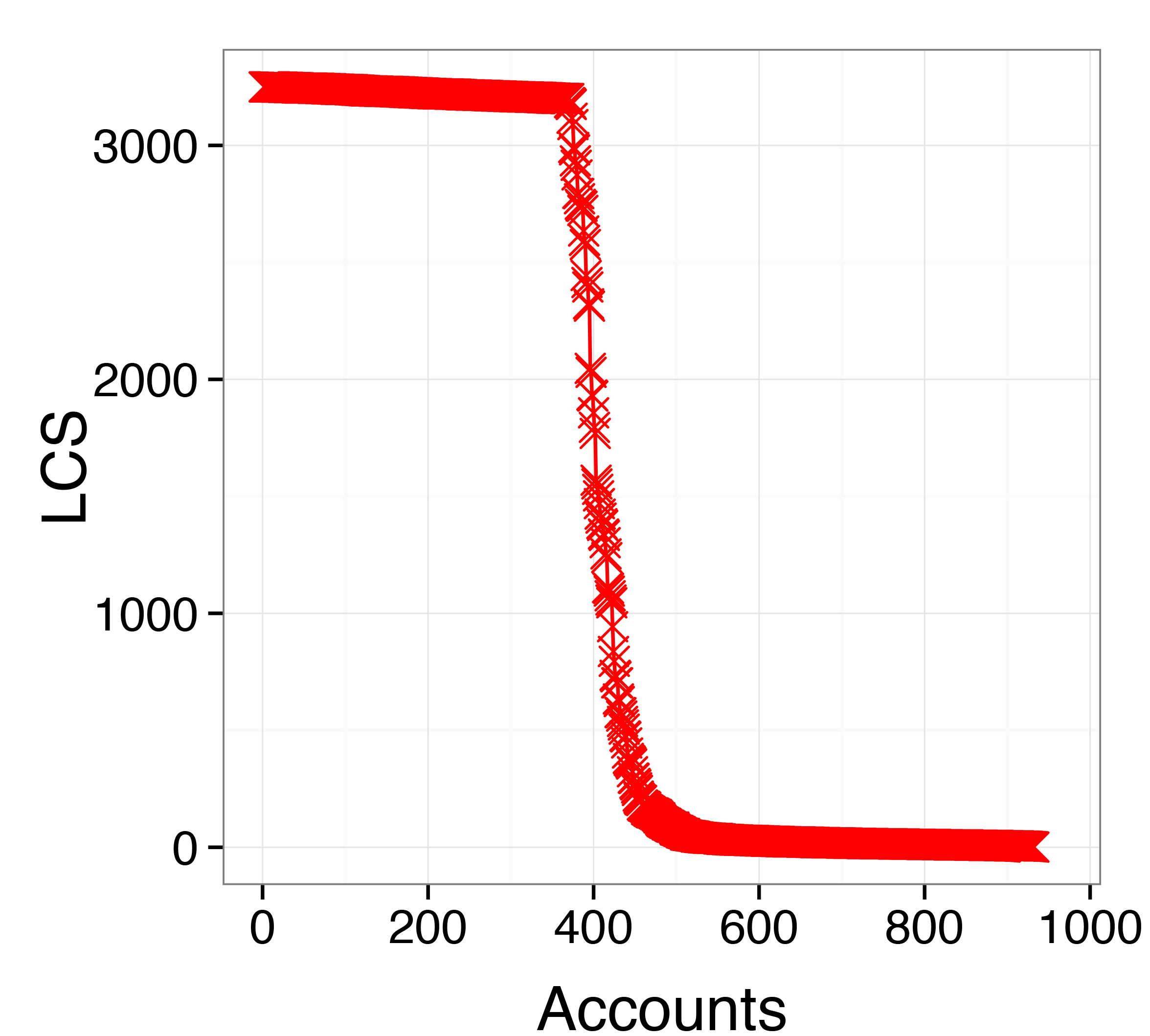}}
  \caption{LCS curves for two groups of heterogeneous accounts, modeled via the \Bttype\ alphabet.
  \label{fig:lcs-mixed}}
\end{figure}

In the above section, we have analyzed LCS curves derived from digital
DNA sequences of users with similar characteristics, such as genuine
Twitter accounts and spambots of a given family. We saw that groups with different characteristics lead to qualitatively
different LCS curves.  However, we have not yet considered LCS curves
obtained from sequences of an unknown and heterogeneous group of
users.  Thus, leveraging the different groups of accounts studied
until now, we built 2 sets of heterogeneous accounts, where we mixed
together all the spambots of the \verb|Bot1| and \verb|Bot2| groups,
with an equal number of genuine accounts. Henceforth, such
heterogeneous groups of accounts are referred to as \verb|Mixed1| and
\verb|Mixed2|, respectively. Figure~\ref{fig:lcs-mixed} shows the LCS
curves obtained via the $\mathbb{B}^3_{\mathit{type}}$ alphabet. 

In the left hand plot of Figure~\ref{fig:lcs-mixed}, we observe a
continuous decrease in the LCS length as the number of considered
accounts grows.  Such slow decrease is sometimes interleaved by
steeper drops, such as those occurring in the region of 500 and 1,000
accounts. Another -- and even more evident -- steep drop is shown in the
right hand plot of Figure~\ref{fig:lcs-mixed}, in the region of 400
accounts. LCS curves in both plots asymptotically reach their minimum
value as the number of accounts grows.  Overall, such LCS curves show
a different behavior than those related to a single group of similar
accounts, such as the ones shown in Figures~\ref{fig:lcs-hum-vs-bot1}
and~\ref{fig:lcs-hum-vs-bot3}. Indeed, the plots of
Figure~\ref{fig:lcs-mixed} lack a single trend that spans for the
whole domain of the LCS curves. Instead, they depict a situation where
a trend seems to be dominant only until reaching a certain
threshold. Then, a steep fall occurs and another -- possibly
different -- trend kicks in. Notably, such portions of the LCS curves
separated by the steep drops resemble LCS curves of the single groups
of similar users (i.e., \verb|Bot1|, \verb|Bot2|, human) used to
obtain the sets of heterogeneous users (i.e., \verb|Mixed1|,
\verb|Mixed2|).  The steep drops of LCS curves separate
areas where the length of the LCS remains practically unchanged, even
for significantly different numbers of considered accounts.  In the
left hand plot of Figure~\ref{fig:lcs-mixed}, for instance, the LCS
remains almost unchanged when considering a number of accounts between
500 and 1,000. The same also applies to the right hand plot of
Figure~\ref{fig:lcs-mixed}, for a number of accounts lower than 400.
Such \textit{plateaux} in LCS curves are strictly related to
homogeneous groups of highly similar accounts.  Note that it is
possible to observe multiple plateaux in a single LCS curve, as in the
case of Figure~\ref{fig:lcs-mixed1}.  This represents a situation
where multiple (sub-)groups exist among the whole set of considered
accounts.  Furthermore, the steeper and the more pronounced is a drop in
a LCS curve, the more different are the two subgroups of accounts
split by that drop.

To summarize, LCS curves of an unknown and heterogeneous group of
users can present one or more \textit{plateaux}, which are related to
subgroups of homogeneous (i.e., with highly similar behaviors)
users. Conversely, \textit{steep drops} represent points marking big
differences between distinct subgroups. Finally, \textit{slow and
  gradual decreases} in LCS curves represent areas of uncertainty,
where it might be difficult to make strong hypotheses about the
characteristics of the underlying accounts. In conclusion, we argue
that LCS curves of an unknown and heterogeneous group of users are
capable of conveying information about relevant and homogeneous
subgroups of highly similar users.

\section{Social Fingerprinting: Leveraging LCS curves to detect social spambots}
\label{sec:best-cut-choice}
In this section, we report the results of our experiments performed to
detect the two subgroups of spambots and genuine accounts that
constitute our \verb|Mixed1| and \verb|Mixed2| groups. We discuss two
different methods to split the accounts of the \verb|Mixed1| and
\verb|Mixed2| groups, according to the characteristics of their LCS
curves. We define a supervised and an unsupervised approach, showing the
suitability of LCS curves and also the effectiveness of the detection
mechanisms. With both the approaches, we consider as spambots those
accounts that are related to high LCS values, namely sharing long
behavioral patterns. Conversely, we consider as genuine users those
accounts that share little portions of their digital DNA. Our
methodologies provide a rigorous assessment of the possibility to
detect spambots (namely, subgroups of accounts with very similar
behavior) using the LCS curves of groups of heterogeneous accounts.

\subsection{Finding subgroups of similar users: a supervised approach}
\label{sec:supervised-approach}
In the spambot detection scenario, supervised approaches are commonly
employed to discriminate between spambots and genuine
users. Supervised classifiers start analyzing a \textit{training-set},
where the class of every user is specified (i.e., users are labeled
either as spambots or genuine ones), in order to understand the characteristics
of the two classes of users. Then, they exploit such learned
characteristics to automatically discriminate between spambots and
genuine users in a new set of unlabeled users. In addition, a
\textit{test-set} is used to evaluate and compare the
effectiveness of different classifiers. This approach is typically
performed in all kinds of machine learning classification tasks. 

We devised a methodology to combine LCS curves and user labels
available from a training-set as a supervised approach for the
detection of subgroups of users (i.e., spambots). A good division of
the original set of users into several subgroups is one where all the
users belonging to a given class are assigned to the same subgroup.
In theory, any point of the LCS curve of a heterogeneous group of
users can be used as a splitting point to obtain two subgroups of more
homogeneous users. Intuitively, however, not all possible splitting
points lead to accurate subgroup partitioning.  Using the given
labels, we can evaluate every possible splitting point in the LCS
curve of the training-set users and find the one that yields the best
possible subgroup division. To this regard, every point generates a
different classifier that can be evaluated in terms of machine
learning performance metrics. The LCS value associated to the
classifier that achieves the best results, according to a given metric
of choice, is then used as a threshold to classify users of the
test-set. The different classifiers can also be qualitatively
evaluated by means of ROC curves, where the best classifiers are those
that lay near to the top-left corner of the
plot~\cite{fawcett2006}. The diagonal line in ROC curves is instead
related to a random classifier.

\begin{figure}[t]
  \centering
  \subfigure[\texttt{Mixed1} group.
  \label{fig:lcs-roc-testset1}]{\includegraphics[width=0.24\textwidth]{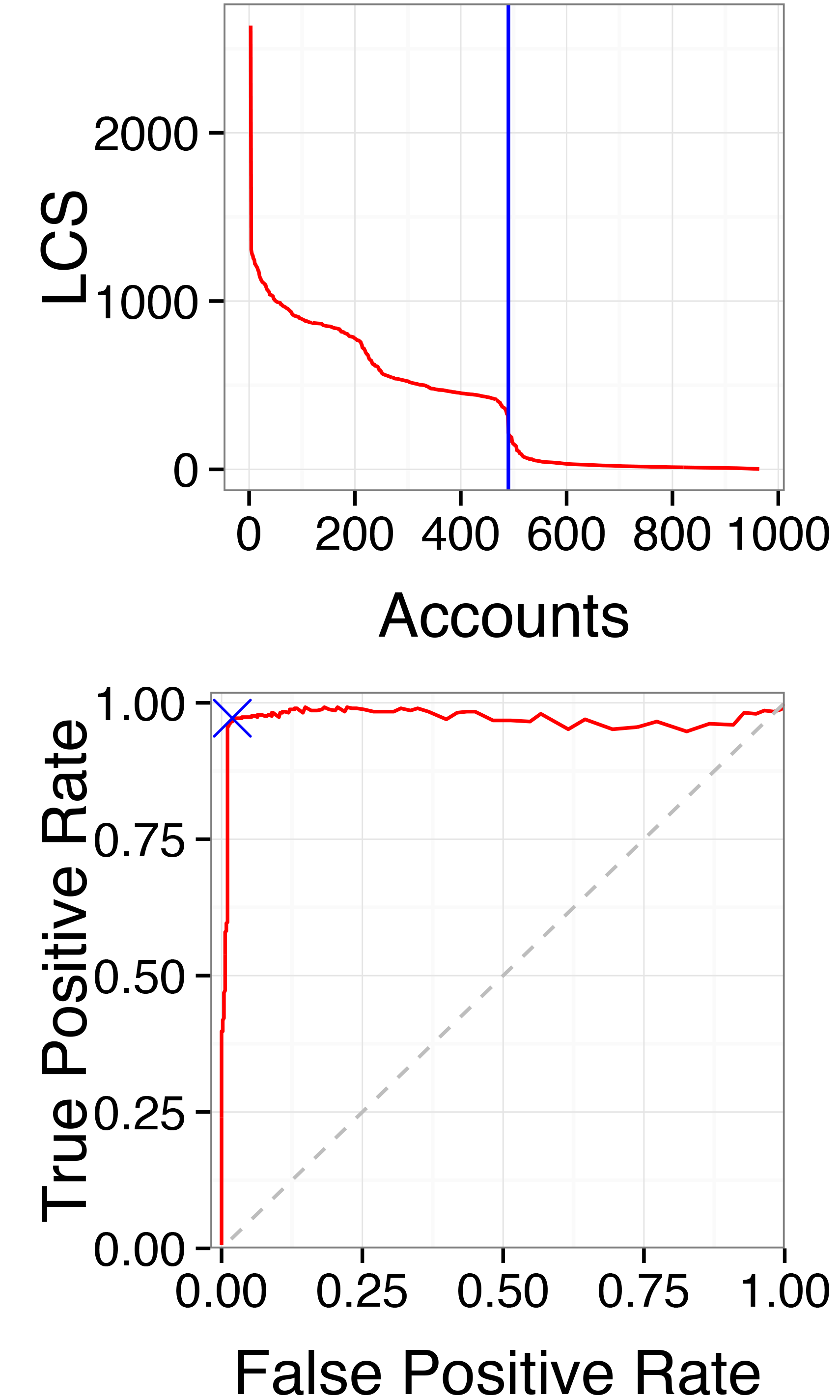}}
  \subfigure[\texttt{Mixed2} group.
  \label{fig:lcs-roc-testset3}]{\includegraphics[width=0.24\textwidth]{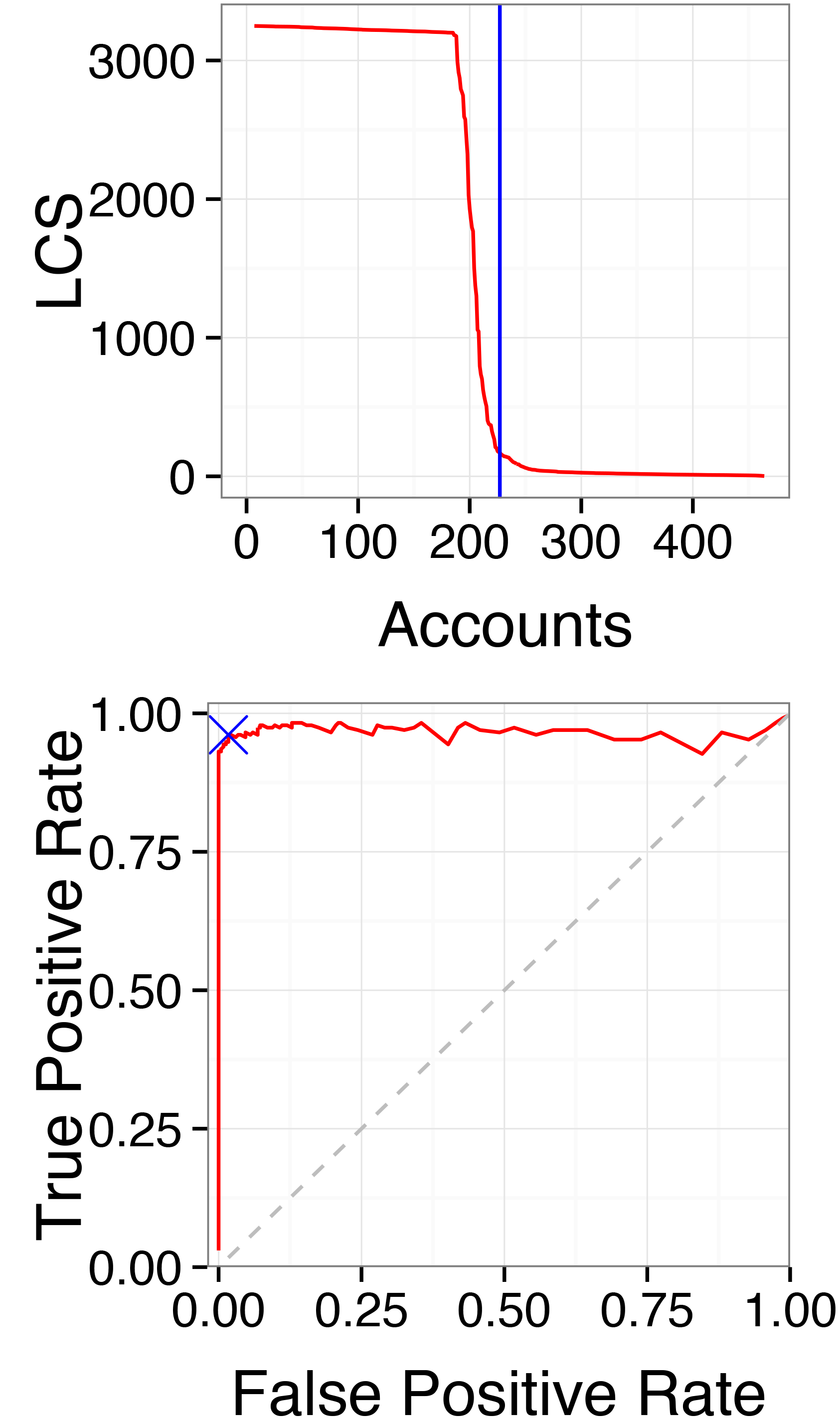}}
  \caption{Application of the supervised approach for discriminating between spambots and genuine users among an unknown set of accounts. ROC curves of the classifiers obtained with the supervised approach are shown in the bottom plots. For each group of users, the best classifier is denoted by a blue cross mark in the corresponding ROC curve. The LCS curves of the 2 heterogeneous groups are shown in the top plots. The best splitting points are marked in the LCS plots with a vertical solid blue line. Accounts to the left of the splitting points are identified as spambots while those to the right are identified as genuine users.
  \label{fig:bestcut-supervised}}\vspace{-0.4cm}
\end{figure}

We evaluated the effectiveness of the aforementioned supervised methodology for the detection of spambots among the \verb|Mixed1| and
\verb|Mixed2| groups of heterogeneous users introduced in
Section~\ref{sec:heterogeneous-lcs}. We used 50\% of \verb|Mixed1|
and \verb|Mixed2| users as the training-set and the remaining part as
the test-set. Figure~{\ref{fig:bestcut-supervised}} shows the LCS
curves of the training-set users and the ROC curves of the respective
classifiers. Among the various metrics commonly adopted to evaluate
machine learning classifiers, we picked the best classifier as the one
achieving the highest \textit{Matthews Correlation
  Coefficient (MCC)}\footnote{The definition and the meaning of the
  \textit{MCC} evaluation metric are given in the following
  Section~\ref{sec:evaluation-comparison}.}~\cite{baldi2000}. In the ROC curves of
Figure~{\ref{fig:bestcut-supervised}}, the points corresponding to the
best classifiers are highlighted with a blue cross mark. The LCS
values associated to the best classifiers represent the best splitting
points according to our supervised approach and are highlighted in
the LCS curves of Figure~{\ref{fig:bestcut-supervised}} as solid
vertical blue lines. 
We identified those users laying to the left of the vertical splitting line as spambots, and those other users laying to the right of the vertical splitting line as genuine.

\begin{figure*}
  \centering
  \subfigure[\texttt{Mixed1} group.
  \label{fig:derivatives-testset1}]{\includegraphics[width=0.40\textwidth]{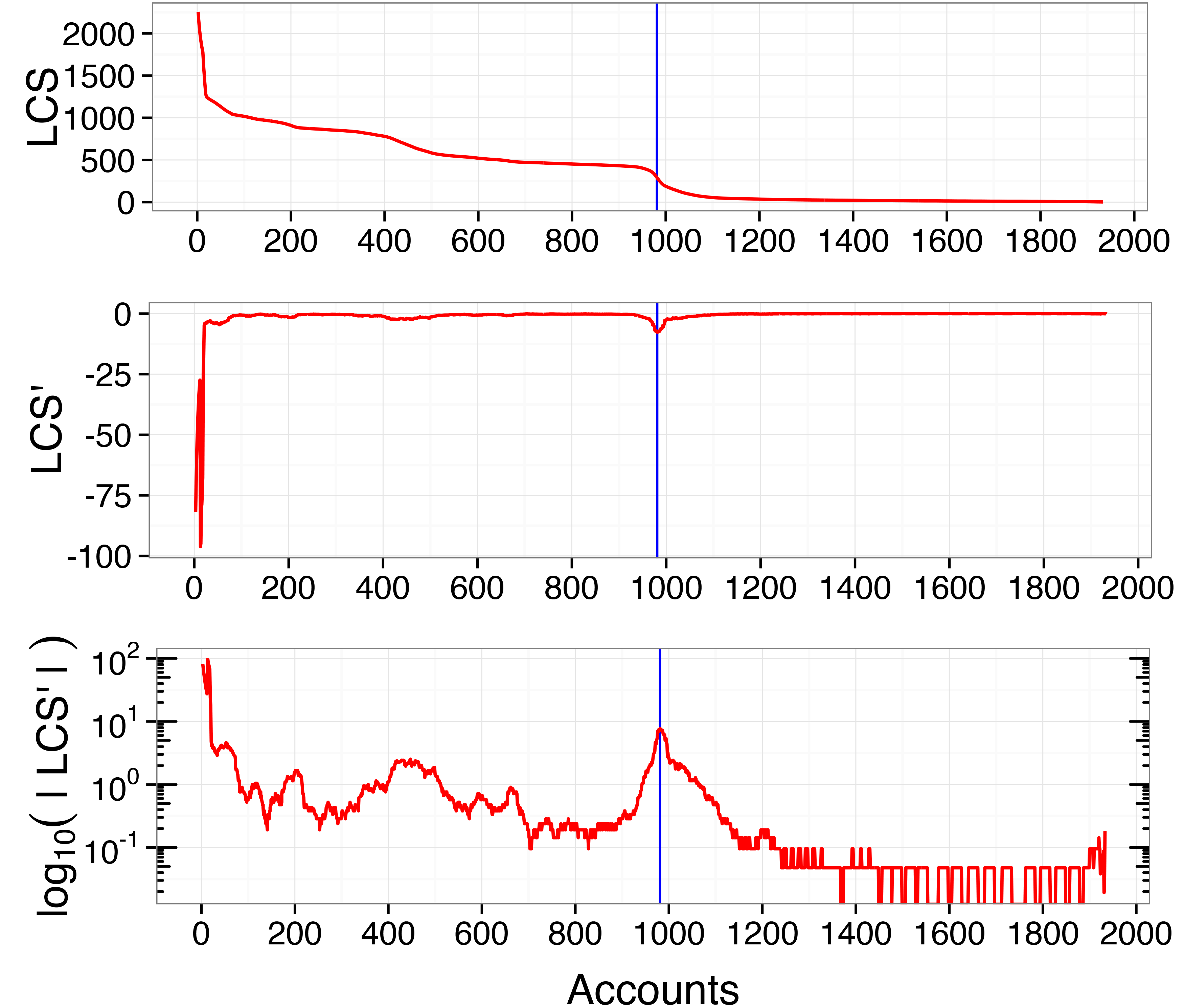}}
  \hspace{1cm}  \subfigure[\texttt{Mixed2} group.
  \label{fig:derivatives-testset3}]{\includegraphics[width=0.40\textwidth]{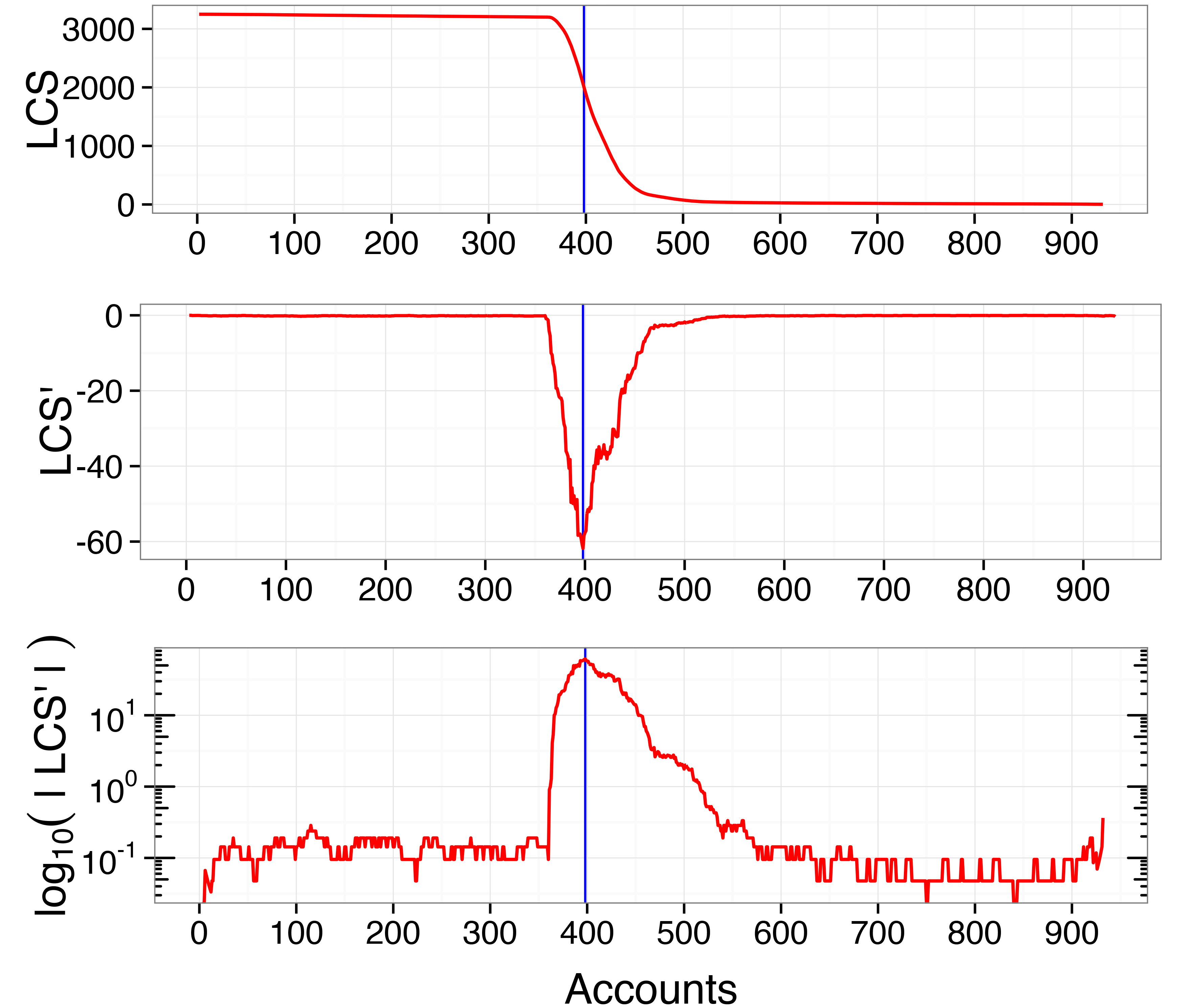}}
  \caption{Application of the unsupervised approach for discriminating between spambots and genuine users among an unknown set of accounts. The peaks in the LCS derivatives that represent the best candidate points for the split are marked in all graphs with a vertical solid blue line. Accounts to the left of the splitting points are identified as spambots, while those to the right are identified as genuine users.
  \label{fig:bestcut-unsupervised}}
\end{figure*}

Notably, as usual for classification approaches that operate in a supervised fashion, one cannot guarantee that the learned LCS value would still be effective when applied on a test-set different from the one used to derive such LCS value.  This problem is known in machine learning literature as \textit{transfer learning} or \textit{inductive learning}~\cite{pan2010}.
In order to overcome this limitation, in the following section we define an unsupervised approach for discriminating between spambots and genuine users, that does not suffer from this drawback.

\subsection{Finding subgroups of similar users: an unsupervised
  approach}
\label{sec:unsup-appr}
Here, we discuss an unsupervised methodology that leverages previous
findings and exploits the shape of LCS curves of heterogeneous users
in order to find subgroups of users with similar behaviors.
Specifically, we propose to exploit the discrete derivative of a LCS
curve to recognize the points corresponding to the steep drops.  This
approach is applicable to a broad range of situations, since it
requires no information other than the LCS curve of the heterogenous
group of users.

The steep drops of LCS curves appear as sharp peaks in the derivative
plot and represent suitable splitting points to isolate different
subgroups among the whole set of
users.  All the suitable splitting points might be ranked according to their
corresponding derivative value (i.e., how steep is the corresponding
drop) and then, a hierarchical top-down (i.e., divisive) approach may
be applied, by repeatedly dividing the whole set of users based on the
ranked points, leading to a dendrogram structure. For instance, this
approach can be exploited in situations where the LCS curve exhibits
multiple plateaux and steep drops, in order to find the best possible
clusters that can be used to divide the original set of heterogeneous
users.

The discrete derivative of the LCS curve of a set of $M$ users can be
easily computed as
\begin{equation*}
\label{eq:lcs-derivative}
\text{LCS}'[k] = \frac{\Delta_k\;\text{LCS}}{\Delta_k\;\text{accounts}} = \frac{\text{LCS}[k] - \text{LCS}[k-1]}{1}
\end{equation*}
for $k = 3, \ldots , M$.
Given that LCS curves are monotonic nonincreasing functions defined
over the $[2,M]$ range, their derivatives LCS$'$ will assume only zero
or negative values, with steep drops in the LCS corresponding to sharp
negative peaks in LCS$'$. Simple peak-detection algorithms can be
employed in order to automatically detect the relevant peaks in
LCS$'$~\cite{palshikar2009}. Notably, this approach does not require a
training phase and can be employed pretty much like a clustering
algorithm, in an unsupervised fashion.

To prove the effectiveness of this unsupervised approach, we applied it
to the LCS obtained from the unlabeled \verb|Mixed1| and \verb|Mixed2|
groups, with the goal of separating spambots from genuine users.
Figures~\ref{fig:derivatives-testset1} and~\ref{fig:derivatives-testset3} show stacked plots of the LCS of
the \verb|Mixed1| and \verb|Mixed2| groups respectively, together with
their discrete derivative LCS$'$, both in linear and logarithmic
scale.  The logarithmic scale plots of the derivatives have been
computed as $\log_{10} |\;\text{LCS}'\;|$ and they have been added for the
sake of clarity, since they highlight the less visible peaks of the
linear scale plots.  In order to facilitate the detection of peaks in
LCS$'$, we smoothed the original LCS curves before computing their
derivatives~\cite{lampos2012}. This preprocessing step acts pretty
much like a low-pass filter, allowing to flatten the majority of noisy
fluctuations.

In Figure~\ref{fig:bestcut-unsupervised}, the solid vertical blue lines correspond to the most
pronounced peaks in the LCS$'$ of \verb|Mixed1| and \verb|Mixed2|. As
shown, the proposed methodology accurately identified reasonable
splitting points in order to find two clusters among the whole sets of
unlabeled users. In detail, those users laying on the left of the vertical
splitting line -- that is, users sharing long behavioral patterns
(i.e., long LCS) -- are labeled as spambots. Conversely, the
users to the right of the vertical splitting line -- i.e., users sharing little
similarities -- are labeled as genuine ones. Together with this qualitative
assessment, in Section~\ref{sec:evaluation-comparison} we also perform
a thorough quantitative evaluation of our spambots detection techniques, by means
of well-known performance metrics of machine learning algorithms.

We remark that, although the \mbox{\texttt{Mixed1} }  and \mbox{\texttt{Mixed2}} groups feature an equal number of genuine and spambot accounts, the ratio between the two types of accounts is generally different when considering the whole Twittersphere\cite{twitters1form}. The balance in \mbox{\texttt{Mixed1} } and \mbox{\texttt{Mixed2}}  is because we mostly envisage the application of our digital DNA technique to spot anomalous groups within devoted events/campaigns, e.g., those accounts retweeting a specific hashtag, or participating in an electoral campaign, or which are followers of a certain account. Whereas the analysis is concentrated on a subset of accounts acting around a particular event, the ratio between genuine and spam accounts can drastically vary, even leading to a balance in the cardinality of the two groups.
For the sake of completeness, we carried out a series of further experiments with the
  aim of investigating the applicability of our technique to the whole
  Twittersphere, in order to gain insights into its effectiveness
  when the ratio between bot accounts and humans is likely different
  than the one in our original test-sets. Results of this experiment are presented in the next section.

\subsection{A comparison of the two approaches}
\label{sec:evaluation-comparison}
As shown in Figure~{\ref{fig:bestcut-supervised}} and Figure~\ref{fig:bestcut-unsupervised}, both the supervised and the
unsupervised approaches identified similar splitting thresholds, that
lay inside the steepest drops of the LCS curves of \verb|Mixed1| and
\verb|Mixed2|. However, results of the supervised and unsupervised
approaches are slightly different, especially with regards to the
accounts of the \verb|Mixed2| group.  In the following, we provide a
quantitative comparison of the two approaches to assess which one
actually better discriminated between spambots and genuine users.

To summarize the outcomes of the supervised and
the unsupervised approaches, we leverage evaluation metrics based on
four standard indicators:
\begin{enumerate}
\item [$\bullet$] {\it True Positives (TP)}: the number of spambots
  correctly recognized;
\item [$\bullet$] {\it True Negatives (TN)}: the number of genuine
  users correctly recognized;
\item [$\bullet$] {\it False Positives (FP)}: the number of genuine
  users erroneously recognized as spambots;
\item [$\bullet$] {\it False Negatives (FN)}: the number of spambots
  erroneously recognized as genuine users.
\end{enumerate}
The meaning of each indicator is summarized by the so-called
\textit{confusion matrix} of Table~\ref{tab:confmatrix}, where each
column represents the instances in the predicted class and each row
represents the instances in the actual (real) class~\cite{kohavi98}:
\definecolor{Gray}{gray}{0.9}
\begin{table}[h!]
\begin{center}
  \footnotesize
    \begin{tabular}{ccc}
      \cellcolor{white} &\multicolumn{2}{c}{\textbf{predicted class}}\\
      \cmidrule{2-3}
      \textbf{actual class} & \textit{genuine user} & \textit{spambot}\\
      \midrule
      \textit{genuine user} & \cellcolor{Gray}\textit{TN} & \cellcolor{Gray}\textit{FP}\\
      \textit{spambot}  & \cellcolor{Gray}\textit{FN} & \cellcolor{Gray}\textit{TP}\\ 
      \midrule
    \end{tabular}
\end{center}
\caption{Confusion matrix.
\label{tab:confmatrix}}
\vspace{-0.5cm}
\end{table}

Then, building on the previously introduced indicators, we computed
the following standard evaluation metrics:
\begin{enumerate}
\item [$\bullet$]\textit{Precision}, the ratio of predicted positive
  cases (i.e., spambots) that are indeed real positives:
  $\frac{\textit{TP}}{\textit{TP+FP}}$;
\item [$\bullet$]\textit{Recall} (or also \textit{Sensitivity}), the
  ratio of real positive cases that are indeed predicted as positives:
  $\frac{\textit{TP}}{\textit{TP+FN}}$;
\item [$\bullet$]\textit{Specificity}, the ratio of real negative
  cases (i.e., genuine users) that are correctly identified as negative:
  $\frac{\textit{TN}}{\textit{TN+FP}}$;
\item [$\bullet$]\textit{Accuracy}, the ratio of correctly classified
  users (both positives and negatives) among all the users:
  $\frac{\textit{TP}+\textit{TN}}{\textit{TP+TN+FP+FN}}$;
\item [$\bullet$]\textit{F-Measure}, the harmonic mean of
  \textit{Precision} and \textit{Recall}:
  $2\cdot\frac{\textit{Precision}\;\cdot\;\textit{Recall}}{\textit{Precision+Recall}}$;
\item [$\bullet$]\textit{Matthews Correlation Coefficient}
  (\textit{MCC})~\cite{baldi2000}, the estimator of the correlation
  between the predicted class and the real class of the users:\\
  \(
  \frac{TP \cdot TN - FP \cdot FN}{\sqrt{(TP+FN) \cdot (TP+FP) \cdot
      (TN+FP) \cdot (TN+FN)}}
  \)
\end{enumerate}
\makeatletter{}\newlength{\tabrowskip}
\setlength{\tabrowskip}{0.5ex}
\begin{table*}[t]
	\footnotesize
	\centering
	\begin{tabular}{llc@{\phantom{M}}rrrrc@{\phantom{M}}rrrrcr}
		\toprule
		&&& \multicolumn{4}{c}{\textbf{detection results}} && \multicolumn{6}{c}{\textbf{evaluation metrics}} \\
		\cmidrule{4-7} \cmidrule{9-14}
		\textbf{} & \textbf{type} && \textit{TP} & \textit{TN} & \textit{FP} & \textit{FN} && \textit{Precision} & \textit{Recall} & \textit{Specificity} & \textit{Accuracy} & \textit{F-Measure} & \textit{MCC}\\
		\midrule
		\multicolumn{14}{l}{\texttt{Mixed1}} \\ [\tabrowskip]
			Social  Fingerprinting					& unsupervised	&& 963	& 924	& 18		& 28		&& \textbf{0.982}	& 0.972	& \textbf{0.981}	& 0.976	& \textbf{0.977}	& 0.952 \\ [\tabrowskip]
			Social  Fingerprinting					& supervised	&& 965	& 924	& 18		& 26		&& \textbf{0.982}	& \textbf{0.977}	& \textbf{0.981}	& \textbf{0.977}	& \textbf{0.977}	& \textbf{0.955} \\  [\tabrowskip]
			Yang \textit{et al.}~\cite{yang2013}		& supervised	&& 169	& 811	& 131	& 822	&& 0.563	& 0.170	& 0.860	& 0.506	& 0.261	& 0.043 \\ [\tabrowskip]
			Miller \textit{et al.}~\cite{miller2014}		& unsupervised	&& 355 	& 657	& 285	& 636	&& 0.555	& 0.358	& 0.698	& 0.526	& 0.435	& 0.059 \\ [\tabrowskip]
			Ahmed \textit{et al.}~\cite{ahmed2013}	& unsupervised	&& 935	& 888	& 54		& 56		&& 0.945	& 0.944	& 0.945	& 0.943	& 0.944	& 0.886 \\
		\midrule
		\multicolumn{14}{l}{\texttt{Mixed2}} \\ [\tabrowskip]
			Social  Fingerprinting					& unsupervised	&& 398	& 468	& 0		& 66		&& \textbf{1.000}	& 0.858	& \textbf{1.000}	& 0.929	& 0.923	& 0.867  \\ [\tabrowskip]
			Social  Fingerprinting					& supervised	&& 446	& 458	& 10		& 18		&& 0.978	& \textbf{0.961}	& 0.979 & \textbf{0.970}	& \textbf{0.970}	& \textbf{0.940} \\  [\tabrowskip]
			Yang \textit{et al.}~\cite{yang2013}		& supervised	&& 190	& 397	& 71		& 274	&& 0.727	& 0.409	& 0.848	& 0.629	& 0.524	& 0.287 \\ [\tabrowskip]
			Miller \textit{et al.}~\cite{miller2014}		& unsupervised	&& 142	& 306	& 162	& 322	&& 0.467	& 0.306	& 0.654	& 0.481	& 0.370	& -0.043 \\ [\tabrowskip]
			Ahmed \textit{et al.}~\cite{ahmed2013} $^{\sharp}$	& unsupervised	&& 428	& 427	& 41		& 30		&& 0.913	& 0.935	& 0.912	& 0.923	& 0.923	& 0.847 \\ 
		\bottomrule
	\end{tabular}
	\caption{Comparison between the Social Fingerprinting splitting techniques and other state-of-the-art algorithms towards the detection of spambots among test-set users of the \texttt{Mixed1} and \texttt{Mixed2} groups. \mbox{$^{\sharp}$: With regards} to the feature set of~\cite{ahmed2013}, a few accounts had null values for all the features thus resulting in the impossibility to apply the clustering algorithm to such accounts.}
	\label{tab:comparison}\vspace{-0.5cm}
\end{table*}

Each of the above metrics captures a different aspect of the
prediction performance. \textit{Accuracy} measures how many users are
correctly classified in both of the classes, but it does not express
whether the positive class is better recognized than the other
one. Moreover, there are situations where some predictive models
perform better than others, even having a lower
accuracy~\cite{powers2011}. A high \textit{Precision} indicates that
many of the users identified as spambots are indeed real spambots, but
it does not give any information about the number of spambots that
have not been identified as such. This information is instead provided
by the \textit{Recall} metric, indeed a low \textit{Recall} means that
many spambots are left undetected.  \textit{Specificity} instead
measures the ability in identifying genuine users as such.  Finally,
\textit{F-Measure} and \textit{MCC} convey in one single value the
overall quality of the prediction, combining the other
metrics. Furthermore, \textit{MCC} is considered the unbiased version
of the \textit{F-Measure}, since it uses all the four elements of the
confusion matrix~\cite{powers2011}. Being a correlation coefficient,
\textit{MCC} $\approx 1$ means that the prediction is very accurate, \textit{MCC} $\approx 0$ means that the prediction is no better than random
guessing, and \textit{MCC} $\approx -1$ means that the prediction is heavily in
disagreement with the real class.
Table~\ref{tab:comparison} shows the results of the evaluation for 
the users of  \verb|Mixed1| and \verb|Mixed2| groups, for all the
considered metrics.

Overall, both the supervised and the unsupervised approaches provide
accurate results towards the detection of spambots among users of the
\verb|Mixed1| and \verb|Mixed2| groups. This is represented by the
high values in all the considered metrics shown in
Table~\ref{tab:comparison}. As anticipated, the main differences
between the two approaches are related to the performances against the
\verb|Mixed2| group. In this situation the unsupervised approach
achieves slightly worse results, with \textit{MCC} = 0.867 in contrast with
\textit{MCC} = 0.949 of the supervised approach. Indeed, the unsupervised
approach performs a rather conservative split which results in
degraded performances. This is demonstrated by \textit{FP} = 0 and \textit{FN} = 33,
meaning that none of the accounts labeled as spambots were instead
genuine, but that some spambots were left undetected. With the
supervised approach, instead, the splitting threshold is chosen in such
a way that it results in \textit{FP} = 8 and \textit{FN} = 4, thus leading to
significantly better performances. In addition, results of the
supervised approach are very consistent between the \verb|Mixed1| and
\verb|Mixed2| groups, with only minimal variations among the
considered evaluation metrics. In conclusion, the additional information exploited in the supervised
approach (i.e., the class label of training-set users) results in
slightly better detections. Nonetheless, also the unsupervised approach
is able to provide overall accurate predictions.

\begin{figure}[t]
	\centering
        \includegraphics[width=0.50\textwidth]{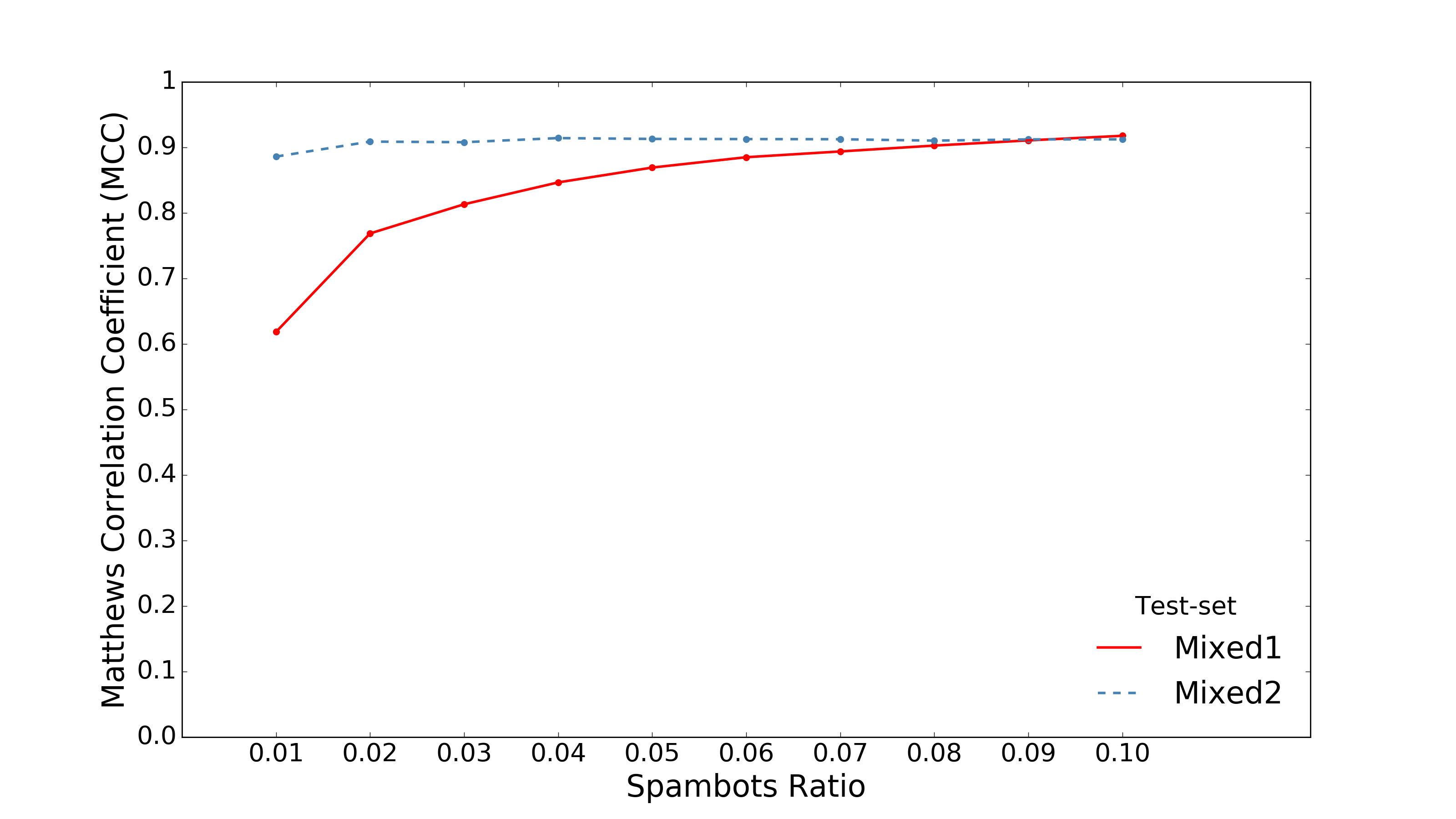}
	\caption{Detection performances of the unsupervised approach with heavily imbalanced data.
	\label{fig:mcc-cuts}}
\end{figure}

Finally, we evaluated the performances of the unsupervised approach with a dataset
  that reflects a real word scenario, where the number of spambots
  is supposed to be much smaller than the number of human-operated
  accounts. In particular, Figure~\ref{fig:mcc-cuts} reports the \textit{MCC} resulted from the experiments obtained considering  the ratio between the number of
  spambots and genuine accounts spanning from 0.01 to 0.10, within a dataset of 5,000 total
  accounts. For each experiment, we firstly set the spambots ratio. Then, we randomly picked the
  DNA sequences of the two original test-sets (\texttt{Mixed1} and
  \texttt{Mixed2}) so as to build mixed datasets with the correct numbers of spambots and genuine accounts. Finally, we executed the
  unsupervised detection approach on such datasets and evaluated the
  detection performance, averaging the results over 20
  executions. From the plot in Figure~\ref{fig:mcc-cuts} it is noticeable that the
  performance improves as the number of bots in the dataset increases. Considering
  that the number of spambot accounts in this experiment is extremely low (the smallest one is
  only of 50 spambots), the reliability of the
  unsupervised approach is still noticeable.
 
\makeatletter{}\section{Discussion}
\label{sec:discussion}
To thoroughly evaluate the Social Fingerprinting technique, we compared our detection results with those obtained by different state-of-the-art  spambot detection techniques, namely the supervised one by Yang \textit{et al.}~{\cite{yang2013}}, and the unsupervised approaches by Miller \textit{et al.}~{\cite{miller2014}} and by Ahmed \textit{et al.}~{\cite{ahmed2013}}.
The work presented in~\cite{yang2013} provides a machine learning classifier that infers whether a Twitter account is genuine or spambot by relying on account's relationships, tweeting timing, and level of automation.
We reproduced such a classifier since the authors kindly provided us with their training set. Instead, works in~{\cite{miller2014}} and~{\cite{ahmed2013}} define a set of machine learning features and apply clustering algorithms. Specifically, in~{\cite{miller2014}} the authors propose modified versions of the DenStream and StreamKM++ algorithms (respectively based on DBSCAN and k-means) and apply them for the detection of spambots over the Twitter stream. Ahmed \textit{et al.}~{\cite{ahmed2013}} exploit the Euclidean distance between feature vectors to build a similarity graph of the accounts and graph clustering and community detection algorithms to identify groups of similar accounts in the graph.

\begin{figure*}[t]
  \centering
  \subfigure[Execution time (\Btcontent ~alphabet).\label{fig:c3-time}]
  {\includegraphics[width=0.22\textwidth]{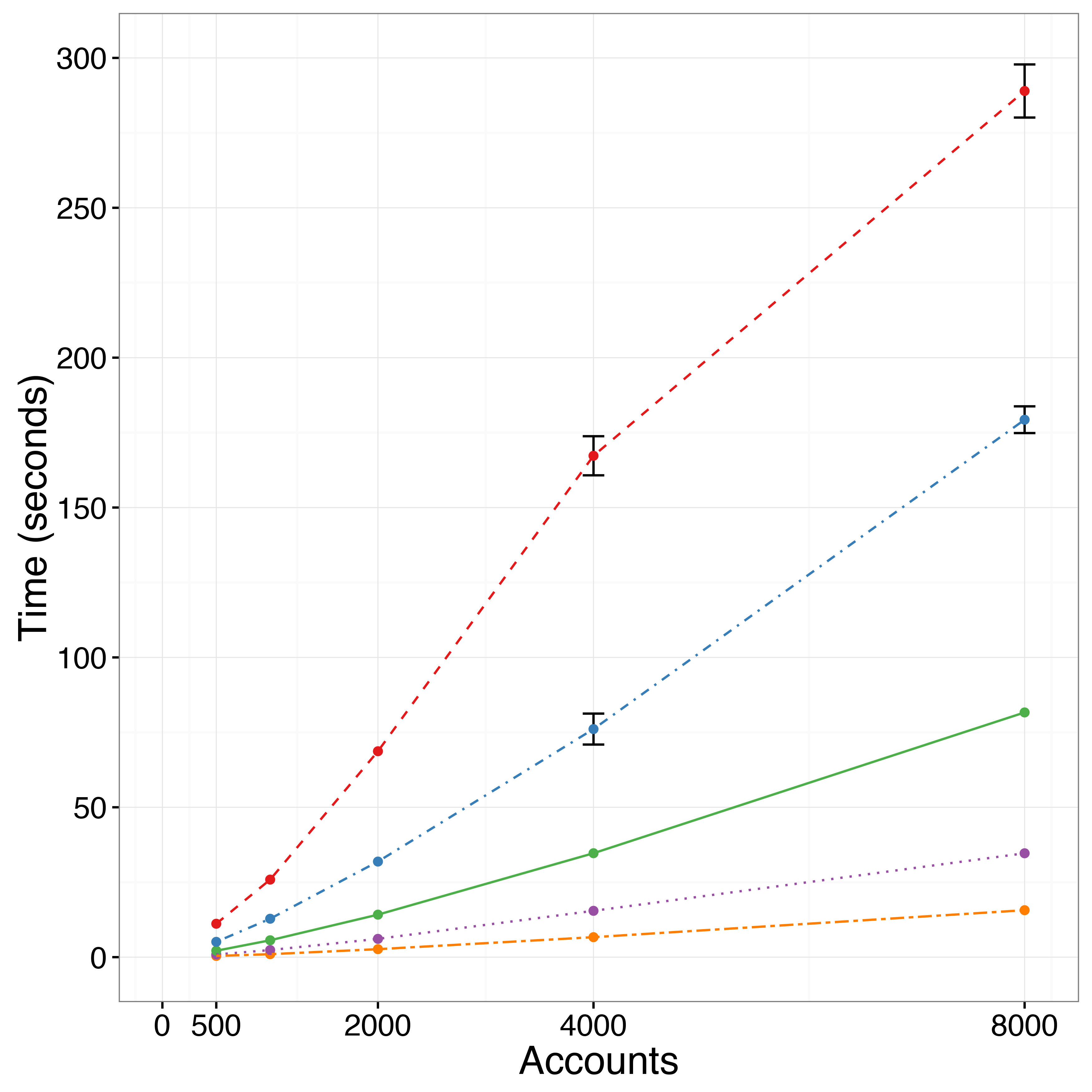}}\hfill
  \subfigure[Memory usage  (\Btcontent ~alphabet).\label{fig:c3-mem}]
  {\includegraphics[width=0.22\textwidth]{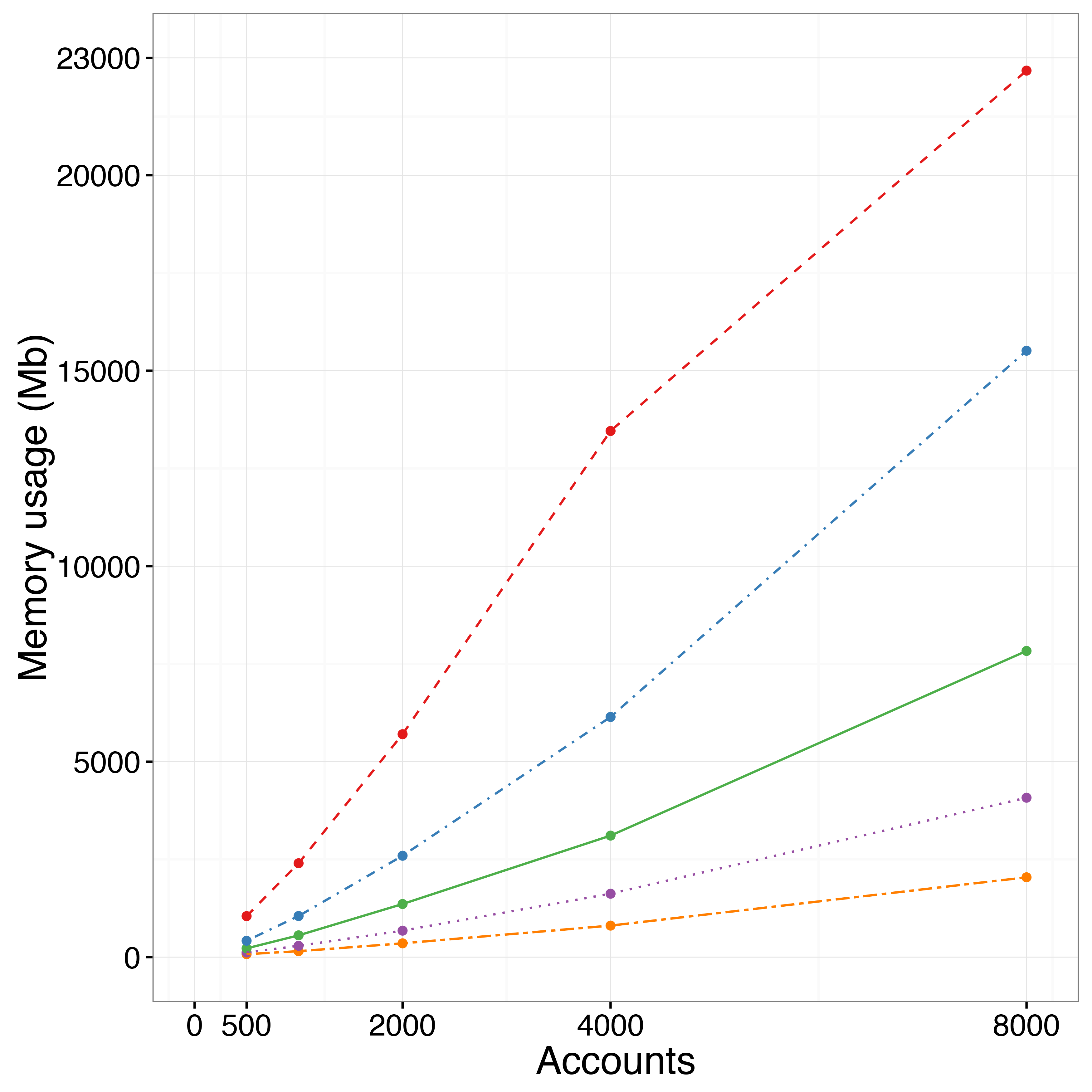}}\hfill
  \subfigure[Execution time (\Bscontent ~alphabet).\label{fig:c6-time}]
  {\includegraphics[width=0.22\textwidth]{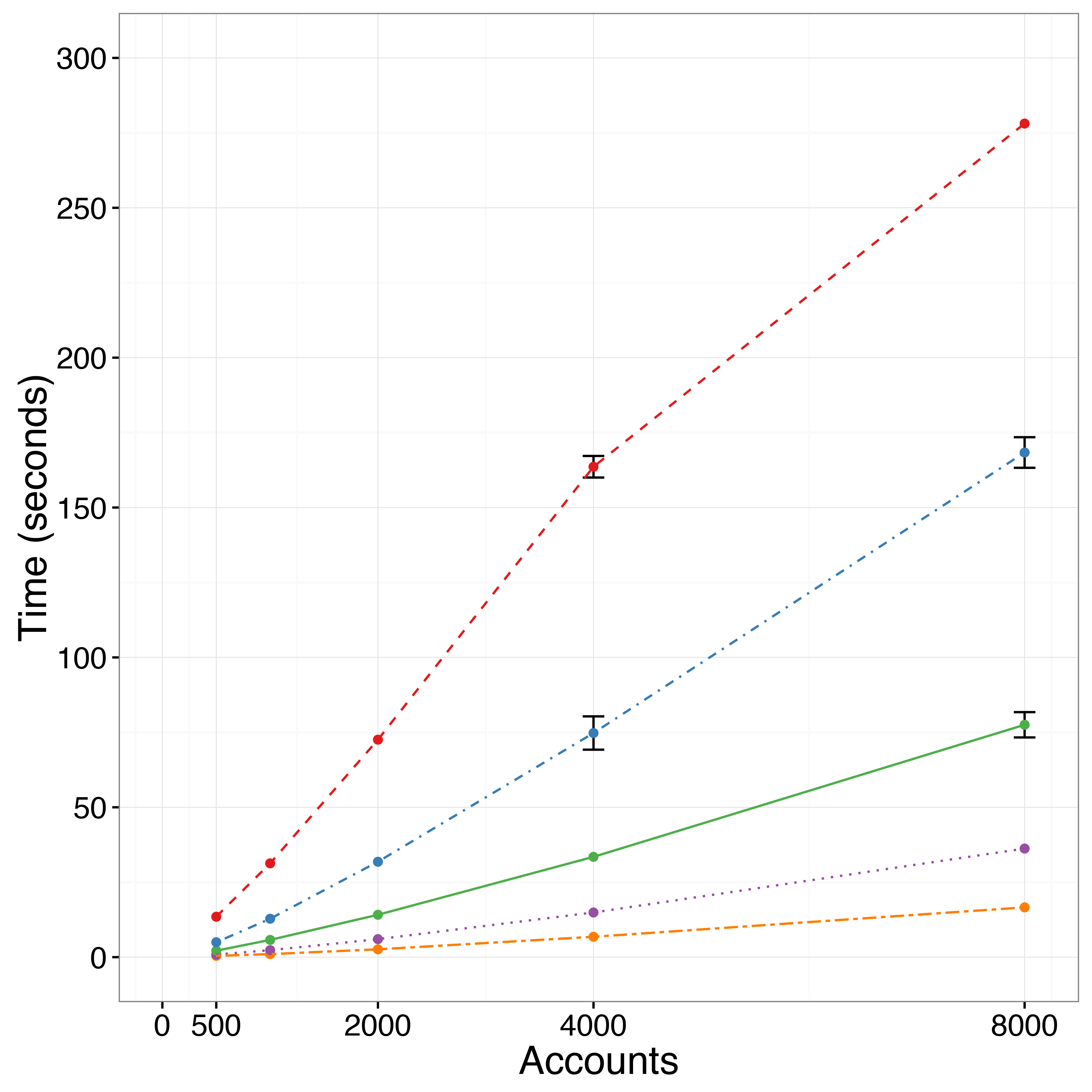}}\hfill
  \subfigure[Memory usage (\Bscontent ~alphabet).\label{fig:c6-mem}]
  {\includegraphics[width=0.28\textwidth]{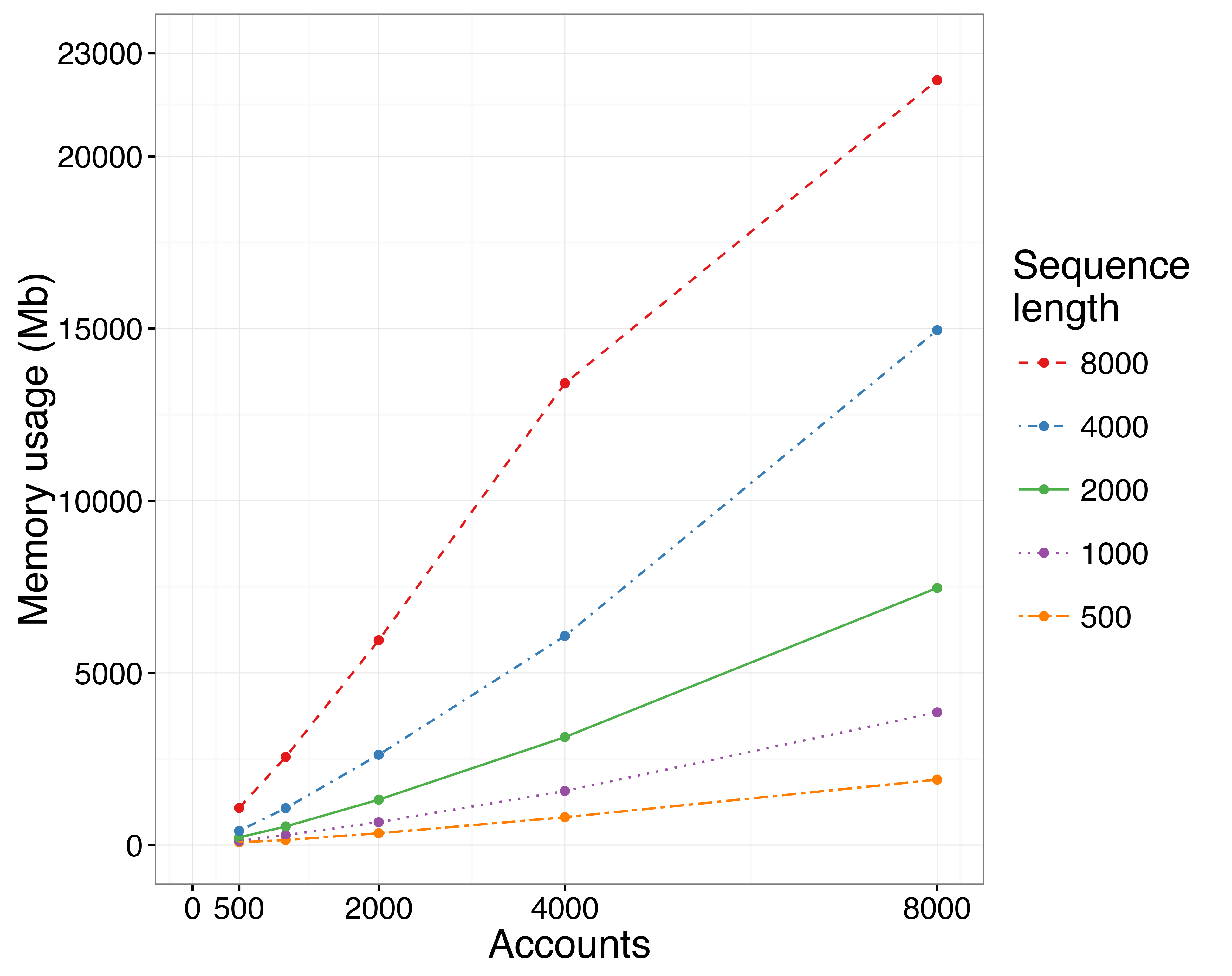}}
  \caption{Mean execution times and memory usages for 20 runs of the Social Fingerprinting detection technique, under different  conditions (i.e., alphabet, number of accounts, length of digital DNA sequences). Error bars report lower/upper bounds of $\pm 3$ standard deviations from the mean.
    \label{fig:complexity}}
\end{figure*}

Notably, the Social Fingerprinting detection technique outperforms the other approaches for all the considered metrics, achieving $\textit{MCC} = 0.952$ and $0.955$ for \texttt{Mixed1}, and $\textit{MCC} = 0.867$ and $0.940$ for \texttt{Mixed2}.
Specifically, there is a clear performance gap between the approaches of~{\cite{yang2013}} and~{\cite{miller2014}} with respect to our proposed approaches and that of~{\cite{ahmed2013}}. The supervised approach by Yang \textit{et al.}~{\cite{yang2013}} proved unable to accurately distinguish spambots from genuine accounts, as demonstrated by the considerable number of false negatives (\textit{FN}) and the resulting very low \textit{Recall}.
This result supports our initial claim that this new wave of bots is surprisingly similar to genuine accounts: they are exceptionally hard to detect if considered one by one.
Moreover, also the unsupervised approach in~{\cite{miller2014}} provided unsatisfactory results. Among the 126 features proposed in~{\cite{miller2014}}, 95 are based on the textual content of tweets. However, novel social spambots, such as those considered in this study, tweet contents similar to those of genuine accounts (e.g., retweets of genuine tweets and popular quotes).
Instead, the approach in~{\cite{ahmed2013}} proved effective in detecting our considered spambots, showing an $\textit{MCC} = 0.886$ for \texttt{Mixed1} and $\textit{MCC} = 0.847$ for \texttt{Mixed2}. With only 7 features, {\cite{ahmed2013}} focuses on retweets, hashtags, mentions and URLs, thus analyzing the accounts along the dimensions exploited by these spammers. However, although achieving an overall good performance for the considered spambots, the approach in~{\cite{ahmed2013}} might lack reusability across other groups of spambots with different behaviors, such as those perpetrating a follower fraud~\cite{jiang2016},\cite{cresci2015}. Social Fingerprinting, instead, is flexible enough to highlight suspicious similarities among groups of accounts without focusing on specific characteristics.

\subsection{Emerging novel spambots}
As introduced in Section~\ref{sec:related-works}, our work tackled the detection of a novel wave of \textit{social Twitter spambots}. By accurately mimicking the characteristics of genuine users, these spambots are intrinsically harder to detect than those studied by Academia in the past years. As a consequence, it is exceptionally difficult to detect such spambots working on an account by account basis, as in the case of machine learning classifiers. This claim is supported by the poor detection results obtained by the approach of Yang \textit{et al.}~\cite{yang2013}. Table~\ref{tab:comparison} clearly shows that the supervised approach of~\cite{yang2013} is unable to effectively distinguish between genuine users and spambots in our 2 mixed datasets. This result is particularly important when considering that the approach of Yang \textit{et al.} was specifically designed to detect \textit{evolving} Twitter spammers. Clearly, such spambots did not evolve in the way that Yang \textit{et al.} imagined. In turn, our work also provides additional evidence of the emergence of a new wave of spambots, as already anecdotally observed in~\cite{ferrara2016}.

Despite the advanced characteristics of these new spambots, we argue that the traces of their automated nature are still present in the history of their behaviors. Such subtle traces might not be enough to infer the nature of an account (whether genuine or spambot) by simply analyzing his past behaviors. Nonetheless, they can be leveraged by observing collective behaviors of groups of accounts. Since spambots of same family -- that is, those spambots belonging to the same bot-master and perpetrating the same illicit activity -- must necessarily have the same goal, and hence similar behaviors, it is possible to exploit behavioral similarities between large groups of accounts as a proxy for automation.
To verify this claim, we have devised the digital DNA modeling technique and we have applied it to the detection of these novel Twitter spambots. Experimental results reported in Section~\ref{sec:best-cut-choice} support our claim and demonstrate the effectiveness of our detection technique. In particular, our supervised and unsupervised Social Fingerprinting approaches outperformed the other detection techniques of Table~\ref{tab:comparison}. Indeed, the proposed technique features the ability to uncover those characteristics that are typical of a group of similar or synchronized accounts. Such characteristics cannot be noticed if the accounts are considered one by one.

\begin{table*}[t]
	\footnotesize
	\centering
	\begin{tabular}{lc@{\phantom{M}}rrrrc@{\phantom{M}}rrrrcr}
		\toprule
		&& \multicolumn{4}{c}{\textbf{detection results}} && \multicolumn{6}{c}{\textbf{evaluation metrics}} \\
		\cmidrule{3-6} \cmidrule{8-13}
		\textbf{} && \textit{TP} & \textit{TN} & \textit{FP} & \textit{FN} && \textit{Precision} & \textit{Recall} & \textit{Specificity} & \textit{Accuracy} & \textit{F-Measure} & \textit{MCC}\\
				\midrule
			Social fingerprinting (\Btcontent)				&& 957	& 894	& 48		& 34		&& 0.952	& 0.966	& 0.949	& 0.958	& 0.959	& 0.915 \\ [\tabrowskip]
			Social fingerprinting (\Bscontent)				&& 958	& 891	& 51		& 33		&& 0.950	& 0.967	& 0.946	& 0.957	& 0.958	& 0.913 \\ [\tabrowskip]
		\bottomrule
	\end{tabular}
	\caption{Detection performances (unsupervised approach) considering the \Btcontent~and \Bscontent~alphabets.}
	\label{tab:performances-b3-vs-b6}\vspace{-0.5cm}
\end{table*}

\subsection{Notes on complexity and scalability}\label{subsec:complexity} Notably, the best performing techniques proposed in recent years for spambots detection are based on data- and time-demanding analyses. This highlights a trade-off between accuracy and responsiveness of spambots detection. The amount of data needed to calculate features -- and the resulting lack of responsiveness in providing results -- also undermine the large-scale applicability of such detection techniques.
To this regard, our Social Fingerprinting technique is not only \textit{effective}, but also \textit{efficient}.
Indeed, some of the previously mentioned approaches for spambots detection are among those requiring a large number of data-demanding features and computationally demanding algorithms. For instance, approaches that are based on graph mining, such as~\cite{jiang2016}, have been proved to be more demanding in terms of data that is needed in order to perform the detection~\cite{cresci2015}. Instead, the Social Fingerprinting technique only exploits Twitter timeline data to perform spambots detection, and executed algorithms run in linear time and disk-space with the number of accounts to investigate~\cite{arnold2011}. In addition, to address and solve other research challenges in the field of social networks, other algorithms for the analysis of biological DNA and strings can be drawn from the established literature in the fields of bioinformatics and string mining. 

To provide experimental evidence of the efficiency and scalability of our detection technique we also conducted some benchmarks, monitoring both time and memory consumption under different experimental settings. In particular, we monitored the effects of increasing the number of investigated accounts, increasing the length of their digital DNA sequences, and changing the considered digital DNA alphabet ({\Btcontent} and {\Bscontent}), on execution time and memory consumption. Results are shown in Figure\mbox{~\ref{fig:complexity}}, reporting mean values of 20 runs that we executed under each experimental setting. Our experiments were performed on a virtual machine with 8Gb of RAM and with a single processor at 2.1Ghz, running Ubuntu.
All the plots in Figure\mbox{~\ref{fig:complexity}} show that the growth is linear for both the time complexity and the required memory, namely doubling one of the parameters roughly doubles the experiment complexity. Considering that the LCS problem is still well studied and has several solutions leveraging high parallelization\mbox{~\cite{chen2006fast}} in distributed computing environments, we can conclude that our approach is scalable enough and that it can be adopted to deal with real cases, e.g., looking for groups of spambots within the followers of a given Twitter account.

\subsection{Flexibility and multidimensionality of digital DNA}
\label{subsec:furtherapplDNA}
Provided the great level of flexibility of digital DNA, we also envision the possibility to exploit results of our Social Fingerprinting technique as a feature in a more complex detection system.
For instance, a hybrid detection system leveraging features derived from the digital DNA analysis and other machine learning features, or a system that simultaneously exploits multiple types of digital DNA.
Indeed, different types of DNA (such as \Bttype, \Btcontent\ and \Bscontent, defined in Section~\ref{sec:twitter-dna}) can be exploited to model different dimensions of user behaviors. Then, results of these models could be used simultaneously in an ensemble or voting system. For instance, when modeled with a given digital DNA alphabet, some accounts might lay in an area of uncertainty of the resulting LCS curve, e.g., a slow and gradual decrease, as explained in Section~\ref{sec:heterogeneous-lcs}. However, the same accounts might instead be unambiguously characterized by the LCS curve obtained with a different digital DNA alphabet. Hence, exploiting multiple alphabets (and then different aspects of the users' behavior) might allow to uncover more characteristics of the accounts under investigation, ultimately leading to better detection results.

\newcommand{\Binteract}{\ensuremath{\mathbb{B}_{\textit{interaction}}}}
\begin{figure}[t]
  \centering
  \subfigure[LCS curves of \textit{permuted} vs. \textit{original} DNA sequences of \texttt{Bot1} accounts.\label{fig:perm-bot1}]
  {\hfill\includegraphics[width=0.24\textwidth]{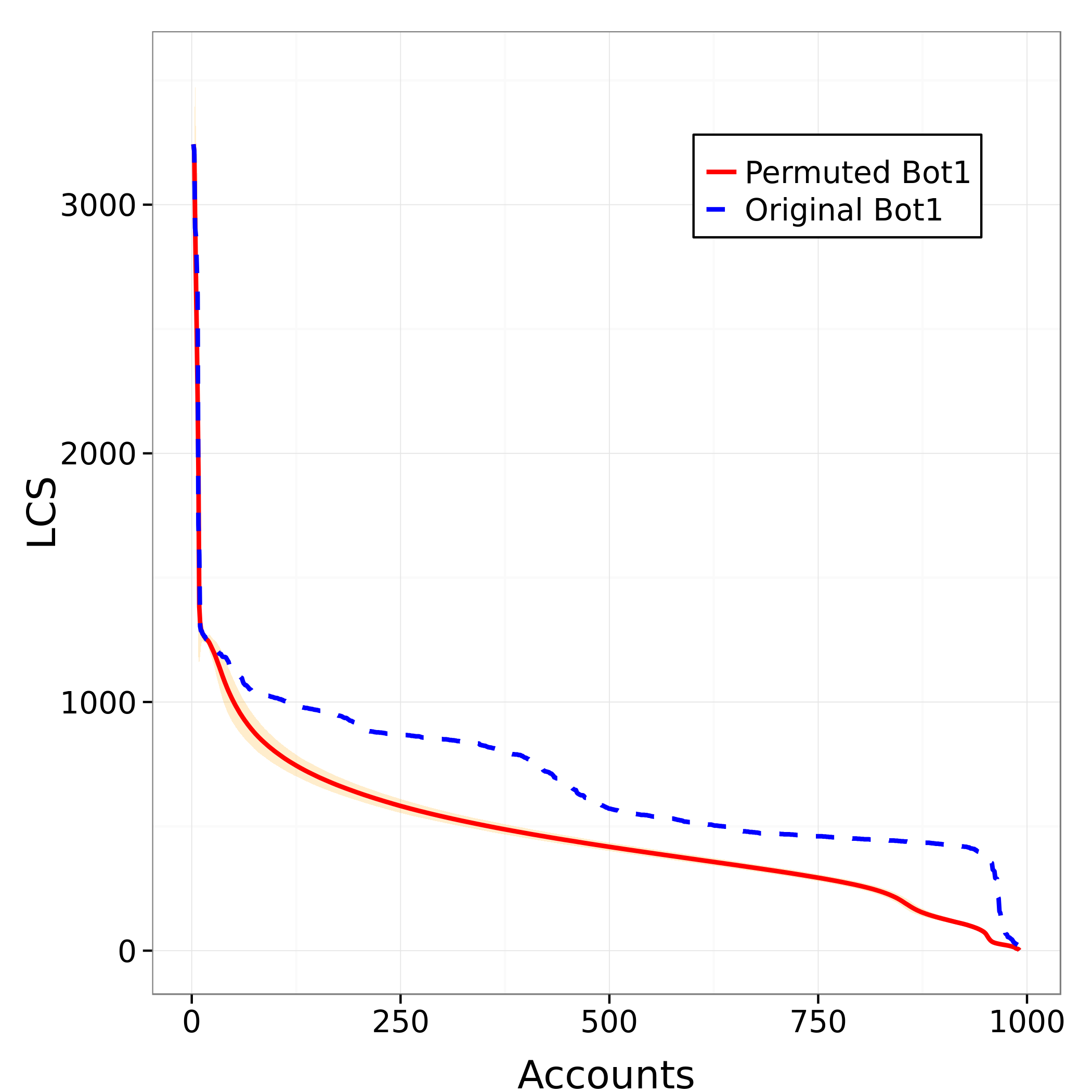}}
  \subfigure[LCS curves of \textit{permuted} vs. \textit{original} DNA sequences of \texttt{Bot2} accounts.\label{fig:perm-bot2}]
  {\includegraphics[width=0.24\textwidth]{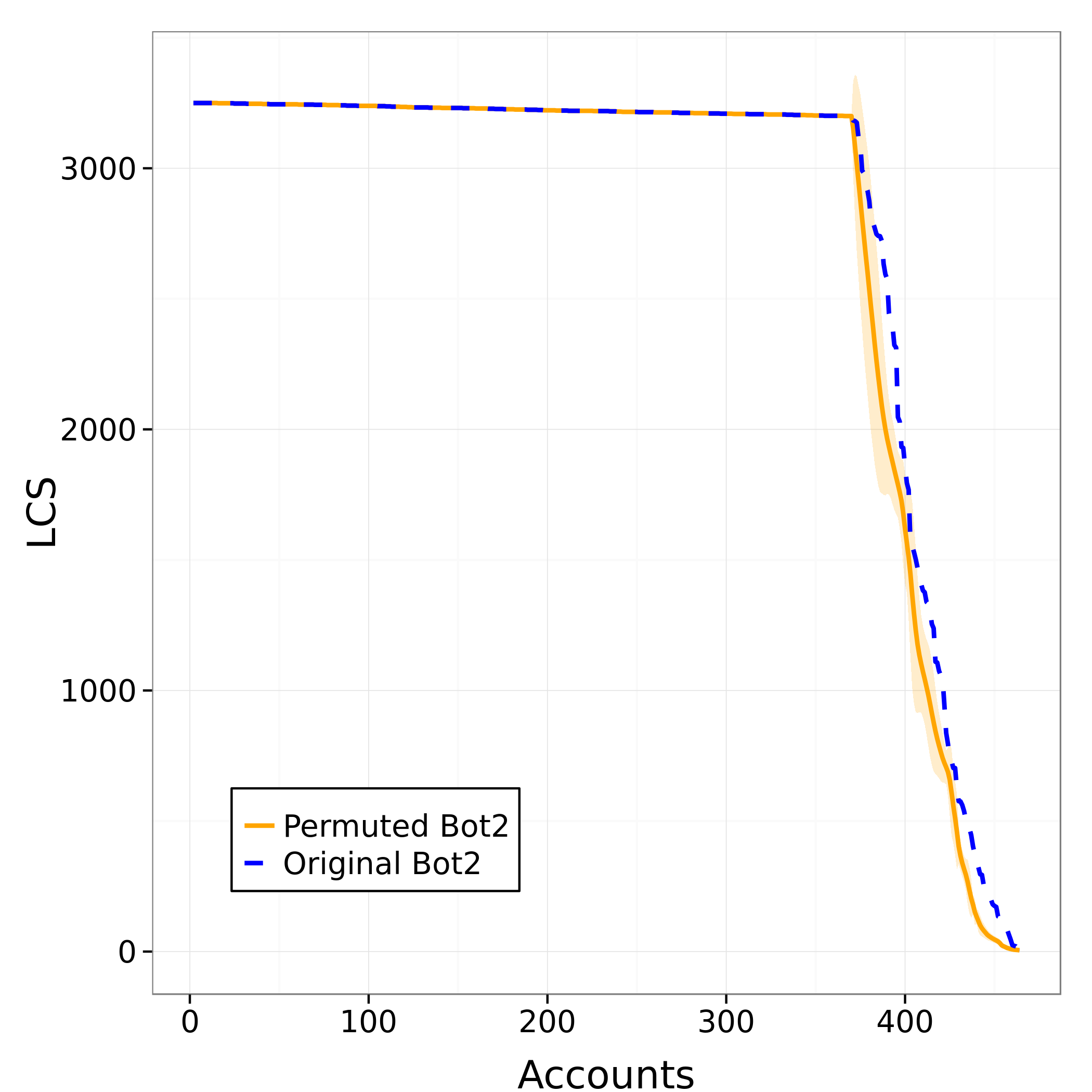}}\\
  \subfigure[LCS curve of genuine (human) accounts and those of permuted \texttt{Bot1} and \texttt{Bot2}.\label{fig:hum-vs-perm}]
  {\includegraphics[width=0.24\textwidth]{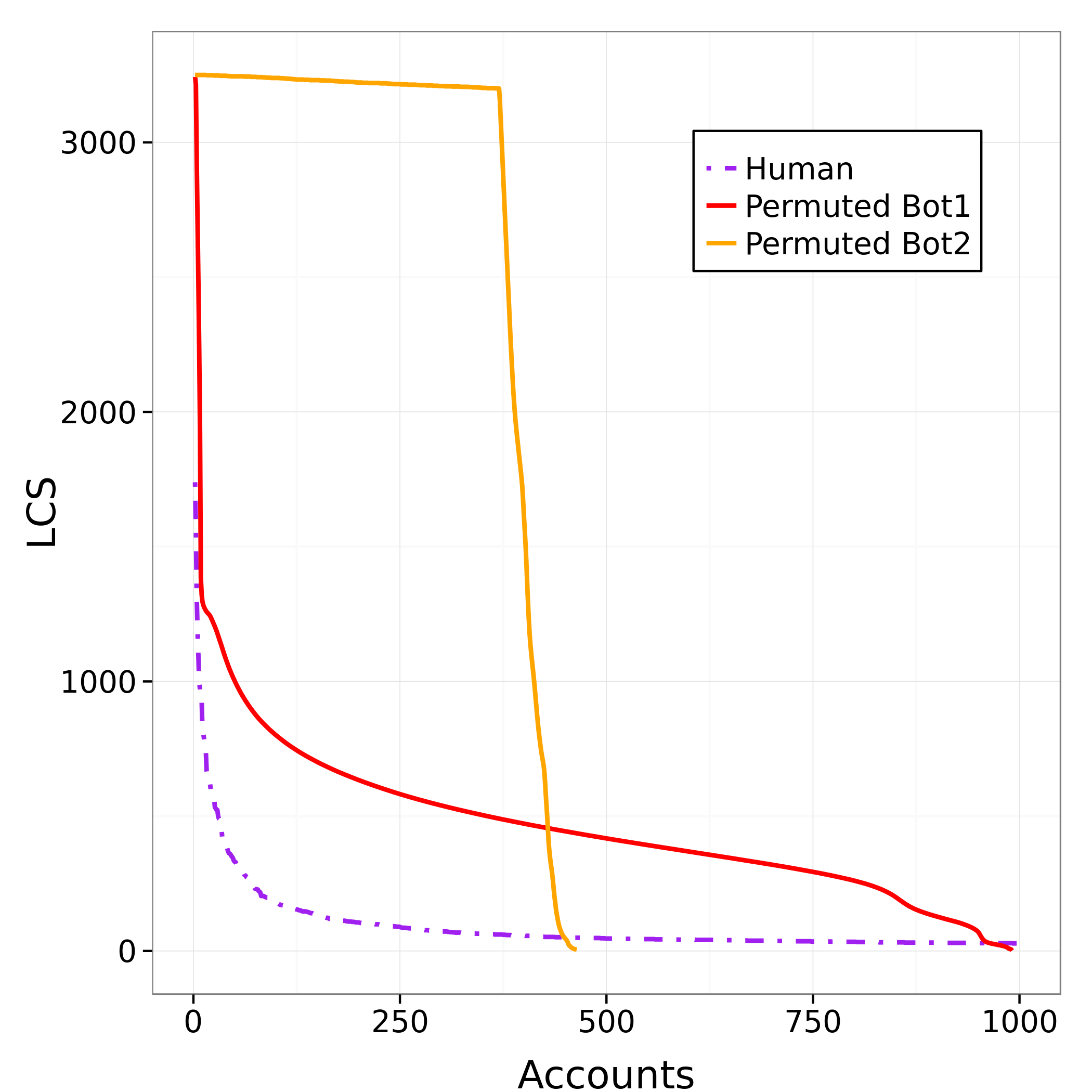}}\hfill
  \caption{Effects of a Monte Carlo simulation with 1,000 random permutations of digital DNA sequences of \texttt{Bot1} and \texttt{Bot2} accounts. LCS curves of permuted sequences report the mean values measured from the 1,000 permutations. Colored areas (ribbons) around the permuted LCS curves highlight lower and upper bounds of $\pm 3$ standard deviations from the mean.
    \label{fig:permutations}}
\end{figure} 

Further alphabets according to which it could be possible to extract the digital DNA of online users are as follows. One alphabet could capture the interaction patterns of Twitter users, considering the popularity level of the peers with
whom a given user interacts.  Specifically, we may think to exploit retweets and
replies among users as a form of interaction and an account's
followers count as a measure of popularity for that account. 
For example,  the alphabet {\Binteract} may define a base to represent interactions with celebrity users, another
 base to represent interactions with ordinary users, and
one last base to represent tweets that are not interactions (i.e.,
that are not retweets nor replies).

It would even be possible
to easily model via our digital DNA sequences both the web-based
online behavioral features recently studied
in~\cite{benevenuto2012,viswanath2014,ruan2016} and the behavioral
features studied in~\cite{cao2010}, derived from the fields
of market analysis and social security analysis.

Experimenting with different alphabets and bases may vary the outcome of our modeling and analysis technique. To evaluate the power of the proposed approach varying the alphabet and the number of bases, Table~\ref{tab:performances-b3-vs-b6} shows the detection performances of Social Fingerprinting -- in its unsupervised fashion -- when using {\Btcontent} and {\Bscontent}, defined in Section~\ref{sec:twitter-dna}, for the sequentialization of the behavior of the \mbox{\texttt{Mixed1}} accounts. Results in Table~\ref{tab:performances-b3-vs-b6} show slightly worse detection performance, with respect to those measured with the {\Bttype} alphabet: \textit{MCC} = 0.915 for {\Btcontent} and \textit{MCC} = 0.913 for {\Bscontent} versus \textit{MCC} = 0.952 for {\Bttype}. Anyway, despite being slightly worse, the detection performance is still extremely good across all evaluation metrics, thus representing the effectiveness and applicability of our technique.

Finally, we notice how our proposed approach is very generic and
flexible, since it makes possible
  to deepen the analysis of LCS curves with the straightforward use of
  powerful tools, such as dendrograms for clustering and ROC curves
  for classifiers, which are already and widely adopted in a number of
  machine learning tasks.

\subsection{Defense against evading techniques}\label{subsec:evading}
Given the focus of the Social Fingerprinting technique on the sequence of user actions, one could foresee evading spammers to randomly re-order the sequence of their tweets, in an effort to escape detection.
In order to thoroughly assess the impact of such evading technique on our detection performances, we have run a series of experiments via Monte Carlo simulations~\cite{kroese2014}. Specifically, for each account of the \texttt{Bot1} and the \texttt{Bot2} groups, we have performed 1,000 random permutations of their real digital DNA strings. Performing a permutation of a digital DNA string literally means randomly changing the order of actions. Then, Figures~\ref{fig:perm-bot1} and~\ref{fig:perm-bot2} show how the LCS changes when applying the permutations. As shown, the values of LCS are lower for the permuted sequences with respect to the original (i.e., real) sequences. This means that randomly re-ordering the sequence of actions actually erased some of the similarities between the spambot accounts. However, the qualitative trend of the LCS curves did not change. In particular, as shown in Figure~\ref{fig:hum-vs-perm}, the LCS curves obtained from the permuted sequences are still significantly different from the long-tailed distribution that is typical of genuine accounts. Such striking difference would still allow to perform a rather accurate detection of the spambots. These spambots are still distinguishable from the genuine accounts, even after the permutation experiment, because the statistical distribution of the bases in the spambot sequences is different with respect to that of the human sequences. On the one hand, spambots have less variability (i.e., entropy) in their sequences and they tend to have a DNA base that is predominant with respect to the other ones. On the other hand, genuine accounts feature a base distribution that is almost uniform. This explains why, in our data, randomly reordering the sequences of spambots, although partly erasing similar behavioral patterns, still allows to distinguish them from humans, according to our LCS similarity approach.

To deal with the spambot accounts under investigation, we relied on the longest common substring metric~\cite{arnold2011}. This metric is rather rigid since it considers exact matches of substrings among two or more digital DNA sequences. Future spambots could make their sequence of actions more random and with higher entropy, thus possibly evading the longest common substring similarity metric. However, we argue that our technique could still be adopted by relying to more flexible similarity metrics on strings. One notable and promising example is the \textit{longest common subsequence} metric~\cite{bergroth2000}. Such metric, already largely adopted in a number of biological DNA analysis tasks, extends the longest common substring by also considering partial matches instead of exact ones. Thus, it could be exploited to uncover an even larger set of behavioral similarities left behind by sophisticated social spambots.

The same metric could also be used for the detection of other -- more sophisticated -- types of spammers, such as \mbox{{\it crowdsourcing spammers}} -- i.e., humans hired to perform spamming tasks. While crowdsourcing spammers may demonstrate a certain level of similarity, they probably manifest behaviors that are not as consistent as the ones of automated accounts (i.e., bots). As such, detection mechanisms targeting crowdsourcing spammers may benefit from leveraging the more flexible longest common subsequence.

\makeatletter{}
\section{Conclusions}
\label{sec:conclusions}
In this paper, we first confirmed that recent waves of spambots have been thoroughly engineered so as to mimic the human behavior of OSNs genuine users. We also proved that these novel species of spambots do escape state-of-the-art algorithms specifically designed to detect them.
Later, we proposed the digital DNA behavioral modeling technique. Leveraging this methodology, we have been able to verify our working hypothesis: there are still low intensity signals that make humans different from bots, when considering users not on an account by account basis, but rather on collective behaviors. 
Our Social Fingerprinting detection approach and coupled algorithmic toolbox -- drawn from the bioinformatics and string mining domains -- have shown excellent detection capabilities for all of the most relevant detection metrics, outperforming state-of-the-art solutions.

\makeatletter{}
\section*{Acknowledgements}
\label{sec:ack}
This research is supported in part by the EU H2020 Program under the
schemes \texttt{INFRAIA-1-2014-2015: Research Infrastructures}
grant agreement \#654024 \textit{SoBigData: Social Mining \& Big
Data Ecosystem} and \texttt{MSCA-ITN-2015-ETN} grant agreement \#675320 \textit{European Network of Excellence in Cybersecurity (NECS)}.
Funding has also been received from the Registro.it IIT-CNR project \textit{MIB (My Information Bubble)}, from Fondazione Cassa di Risparmio di Lucca that partially finances the regional project \textit{Reviewland}, and from the MIUR (Ministero dell'Istruzione, dell'Universit\`{a} e della Ricerca) and Regione Toscana (Tuscany, Italy) funding the \textit{SmartNews: Social sensing for Breaking News} project: \texttt{PAR-FAS 2007-2013}.
\begin{figure}[h]
  \centering
{\includegraphics[width=1\columnwidth]{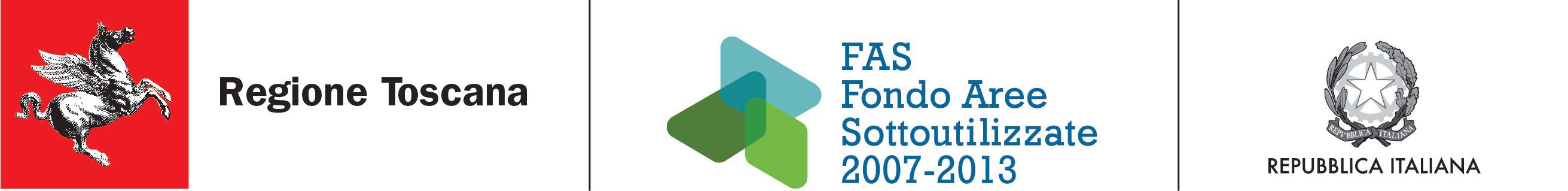}}
\end{figure}

\bibliographystyle{IEEEtran}
\bibliography{references}
\makeatletter{}\section*{Biographical Notes}
\begin{IEEEbiography}[{\includegraphics[width=1in,height=1.25in,clip,keepaspectratio]{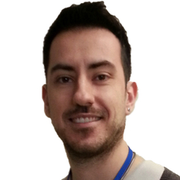}}]{Stefano Cresci}
is a PhD student in the Department of Information Engineering at the University of Pisa and a Research Fellow at the Institute of Informatics and Telematics of the Italian CNR (IIT-CNR).
His research interests include social media mining and knowledge discovery. Cresci received an MSc in computer engineering and an MSc in big data analytics and social mining from the University of Pisa. He is Student Member of IEEE and Member of the IEEE Computer Society.
\end{IEEEbiography}

\begin{IEEEbiography}[{\includegraphics[width=1in,height=1.25in,clip,keepaspectratio]{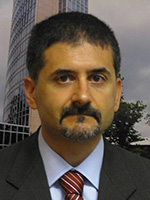}}]{Prof. Dr. Roberto Di Pietro}
  is Global Head Security Research at Nokia Bell Labs, Paris,
  France. His research interests include security, privacy,
  distributed systems, computer forensics, and analytics.  He is also
  Professor at University of Padua, Maths Dept., and affiliated with
  IIT-CNR, Pisa, Italy.
\end{IEEEbiography}

\begin{IEEEbiography}[{\includegraphics[width=1in,height=1.25in,clip,keepaspectratio]{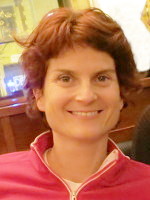}}]{Dr. Marinella Petrocchi}
(M.Sc. 1999, Ph.D.
2005) is a computer science researcher at IIT-CNR, Italy. Her research
interests include privacy and trust aspects
on information exchanged online, including fake accounts and
fake information detection. 
\end{IEEEbiography}

\begin{IEEEbiography}[{\includegraphics[width=1in,height=1.25in,clip,keepaspectratio]{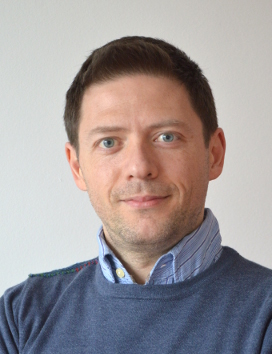}}]{Angelo  Spognardi}
  is Assistant Professor at Compute department of the Technical
  University of Denmark (DTU).  His research interests include network
  security and privacy, security of embedded systems, recommendation
  systems and applied cryptography.
\end{IEEEbiography}

\begin{IEEEbiography}[{\includegraphics[width=1in,height=1.25in,clip,keepaspectratio]{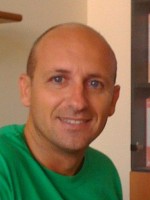}}]{Dr. Maurizio Tesconi}
is a computer science researcher at IIT-CNR. His research interests include social web mining, social network analysis, and visual analytics within the context of open source intelligence. Tesconi received a PhD in information engineering from the University of Pisa. He is a member of the permanent team of the European Laboratory on Big Data Analytics and Social Mining.
\end{IEEEbiography}

\end{document}